\documentclass[preprint,10pt]{elsarticle}
\usepackage{bm}
\usepackage{textcomp}
\usepackage{graphicx}
\usepackage{natbib}
\usepackage{geometry}
\usepackage{pifont}
\usepackage[nodots]{numcompress}
\usepackage{amssymb}
\usepackage{bm} 
\newcommand\md{\,\mathrm{d}}
\journal{Journal of Quantitative Spectroscopy and Radiative Transfer}

\begin{document}
\begin{frontmatter}

\title{An efficient decomposition technique to solve angle-dependent 
Hanle scattering problems}
\author{H. D. Supriya}
\ead{hdsupriya@iiap.res.in}
\author{M. Sampoorna}
\ead{sampoorna@iiap.res.in}
\author{K. N. Nagendra}
\ead{knn@iiap.res.in}
\author{B. Ravindra}
\ead{ravindra@iiap.res.in}
\author{L. S. Anusha}
\ead{anusha@iiap.res.in}

\address{Indian Institute of Astrophysics, Koramangala, Bangalore 560034, 
India}

\begin{abstract}
Hanle scattering is an important diagnostic tool to study weak solar 
magnetic fields. 
Partial frequency redistribution (PRD) is necessary to interpret 
the linear polarization observed in 
strong resonance lines. Usually angle-averaged PRD functions are used to
analyze linear polarization. However it is established that 
angle-dependent PRD functions are often necessary to interpret polarization 
profiles formed in the presence of weak magnetic fields.
Our aim is to present an efficient decomposition technique, and the numerical 
method to solve the concerned angle-dependent line transfer problem.
Together with the standard Stokes decomposition technique we employ 
Fourier expansion over the outgoing azimuth angle to express in a 
more convenient form, the 
angle-dependent PRD function for the Hanle effect. It allows the use of 
angle-dependent frequency domains of Bommier to solve the  
Hanle transfer problem. Such an approach is self-consistent and accurate 
compared to a recent approach where angle-averaged frequency domains were used to 
solve the same problem. 
We show that it is necessary to incorporate angle-dependent frequency domains 
instead of angle-averaged frequency domains to 
solve the Hanle transfer problem accurately, especially for the  
Stokes $U$ parameter. The importance of using angle-dependent domains 
has been highlighted by taking the example of Hanle effect in the case 
of line transfer with vertical magnetic fields in a slab atmosphere. 
We have also studied the case of polarized line formation when 
micro-turbulent magnetic fields are present. The difference between angle-averaged 
and angle-dependent solutions is enhanced by the presence of micro-turbulent 
fields.  
\end{abstract}


\end{frontmatter}
\section{Introduction}
\label{sec_intro}
The polarization of line radiation is caused by resonance scattering on bound 
atomic levels. A modification of this process by external magnetic fields
is called the Hanle effect. The linear polarization of strong resonance lines, 
are particularly sensitive to the type of 
the frequency redistribution mechanism used in their evaluation 
especially in the presence of magnetic fields. 
The differences of the diffuse radiation field between 
the linear polarization ($Q/I$) profiles computed using angle-averaged and 
angle-dependent partial frequency redistribution (PRD) functions are 
illustrated in \citet{mf88} and \citet{sametal11} in the 
non-magnetic (Rayleigh) case. 
\citet{knnetal02} showed that Stokes $U$ profiles computed in 
planar slabs, for the case of Hanle effect, using 
the angle-averaged PRD functions differ significantly from those computed 
using 
angle-dependent PRD functions. \\
\indent In the case of angle-dependent PRD functions, 
the strong coupling that exist between the incoming and scattered radiation 
makes their evaluation and subsequent use in transfer equation 
numerically expensive. The use of decomposition technique developed by 
\citet{hf09} for the Hanle effect and \citet{hf10} for the Rayleigh case, simplifies this 
numerically 
expensive problem. In the non-magnetic case \citet{sametal11} used this 
decomposition technique and 
developed numerical methods to solve the polarized transfer problem 
with angle-dependent PRD functions . 
They also present 
a detailed 
historical account of the works on angle-dependent PRD in spectral line 
polarization. In \citet{sam11b} 
Hanle transfer problem with angle-dependent PRD was solved using single 
scattering approximation. Further in \citet{knnetal11} (hereafter, NS11)
the full Hanle transfer 
problem with angle-dependent 
PRD functions was solved by including multiple scattering terms and using 
scattering expansion method (SEM). 
It may be noted that SEM was first formulated by \citet{hfetal09} for solving the 
polarized line transfer equation with complete frequency redistribution 
(CRD).\\
\indent In all the above mentioned papers, a Fourier-expansion of 
the angle-dependent PRD 
function over azimuth angle difference ($\chi-\chi^\prime$) is employed, 
where $\chi$ and $\chi^\prime$ are the azimuth angles of the outgoing 
and incoming rays. This technique was first introduced by \citet{dh88} and 
was further developed by \citet{hf09} for the Hanle transfer problem. 
The decomposition technique of \citet{hf09} allowed the Hanle transfer 
problem to be solved in an azimuth independent Fourier basis. In 
NS11 this decomposition technique was used together with the  
angle-averaged frequency domains (approximation \rm III of 
\citet{vb97}) to solve the Hanle transfer problem with 
angle-dependent PRD functions. As one 
has to use in principle, angle-dependent domains themselves for 
angle-dependent PRD transfer problems, the approach taken in NS11 is inconsistent. 
Such an approximate approach 
was taken, because, it allowed to work in an azimuth independent Fourier basis. 
Clearly in an axisymmetric Fourier basis one cannot 
apply angle-dependent frequency domains as they explicitly depend on 
azimuth difference ($\chi-\chi^\prime$). This inconsistency is at the 
base of the slight differences in the angle-dependent Hanle solutions 
presented in NS11 (see Fig.4 in that paper), and those presented in \citet{knnetal02}. Indeed 
\citet{lsaknn12} pointed out that the use of angle-averaged frequency 
domains for the angle-dependent Hanle transfer problem (as done in NS11) 
results in a loss of information.\\
\indent To overcome this inconsistency we adopt Fourier expansion of 
angle-dependent Hanle PRD matrix over only the outgoing azimuth angle $\chi$ as 
suggested in \citet{lsaknn12} (see also \citet{lsaknn11}). 
Such an expansion was proposed to solve polarized 
transfer problems in multi-dimensional media, where the radiation field 
is non-axisymmetric even in the absence of a magnetic field. In this paper 
we apply the decomposition proposed by them to the simpler case of 
polarized transfer in 1D media, and in the presence of a magnetic field. We show 
that   
an expansion only over $\chi$ allows to incorporate angle-dependent 
frequency domains for angle-dependent PRD functions `self consistently' 
to solve the transfer problem in the magnetic case. \\
\indent In Sect. 2 of the paper we describe the decomposition technique employed for the 
Hanle effect with angle-dependent PRD. In Sect. 3, we discuss the behavior 
of azimuthal Fourier components of redistribution matrix elements. 
In Sect. 4, we give the equations 
of the SEM to solve the Hanle transfer problem. In Sect. 5, we discuss the 
results obtained by considering our new method of azimuth expansion. 
A comparison  with the results 
obtained 
from the perturbation method described in \citet{knnetal02} and those 
obtained by using angle-averaged domains (in NS11) is done. In the same 
section we also revisit the well known problem of vertical field Hanle effect 
which arises only due to the angle-dependent PRD in line scattering. Further, 
we discuss in detail the role of micro-turbulent magnetic fields on line transfer using 
angle-averaged (AA) and angle-dependent (AD) versions of the redistribution matrix. 
Conclusions are presented in Sect. 6.

\section{The decomposition technique}
\label{decomp}
The polarized transfer equation for the Stokes vector can be written 
in the component form as 
 \begin{equation}
\mu\, {\partial I_i \over \partial \tau} = \left[\varphi(x)+r\right]
\left[I_i(\tau,x,{\bm\Omega})-S_i(\tau,x,{\bm\Omega})\right], 
\label{rte_stokes_component}
\end{equation}
where $i=0,1,2$ refer to the Stokes parameters $(I, Q, U)$ respectively. 
The ray direction is given by $\bm{\Omega} = (\theta, \chi)$, with 
$\theta = \cos^{-1}(\mu) \  \rm{and} \ \chi$ being the polar angles. $x$ is the 
frequency in non-dimensional units. 
The line optical depth 
is denoted by $\tau$ and $\varphi(x)$ is the normalized Voigt function $H(a,\,x)$, 
where $a$ represents a constant damping parameter. 
The ratio of continuum to the line absorption
coefficient is denoted by $r$. The total source vector is given by
\begin{equation}
S_i(\tau,x,{\bm\Omega})={\varphi(x)\,S_{l,i}(\tau,x,{\bm\Omega})+r \,S_{c,i} \over
\varphi(x)+r},
\label{total_source}
\end{equation}
where $S_{c,i}$ are the components of the unpolarized continuum source
vector. We assume that $S_{c,0}=B_{\nu_0}$, where $B_{\nu_0}$ is the Planck 
function at the line center, and $S_{c,1}=S_{c,2}=0$.
The line source vector can be written as
\begin{eqnarray}
&&S_{l,i}(\tau,x,{\bm\Omega})=G_{i}(\tau)
+\int_{-\infty}^{+\infty} \oint \sum_{j=0}^{2}\,
{\hat R_{ij}(x,{\bm \Omega},x^\prime,{\bm \Omega}^\prime, {\bm B})
\over \varphi(x)}\, I_j(\tau,x^\prime,{\bm\Omega}^\prime)\,
{\md\Omega^\prime \over 4\pi}
\md x^\prime,
\label{line_source}
\end{eqnarray}
where ${\bm\Omega}^\prime\ (\theta^\prime,\chi^\prime)$ is the direction of 
the incoming ray defined with respect to the atmospheric normal. 
The 
solid angle element 
$\md \Omega^{\prime} = \sin\theta^{\prime} \md \theta^{\prime} \md \chi^{\prime}$ 
where 
$\theta^{\prime} \in [0,\pi]$ and $\chi^{\prime} \in [0,2\pi]$.
The primary source is assumed to be unpolarized, so that
$G_0(\tau)=\epsilon B_{\nu_0}$ and $G_1(\tau)=G_2(\tau)=0$.
Here, $\hat R_{ij}(x,{\bm \Omega},x^\prime,{\bm \Omega}^\prime, {\bm B})$ is 
the Hanle redistribution matrix with angle-dependent PRD, and $\bm{B}$ represents 
an oriented vector magnetic field. The thermalization parameter 
$\epsilon = \Gamma_I / (\Gamma_R + \Gamma_I)$, with $\Gamma_I$ and $\Gamma_R$ being the 
inelastic collisional de-excitation rate and the radiative de-excitation rate respectively.\\
\indent In the decomposition method used in this paper, 
the Stokes vector $(I, Q, U)$ 
is first decomposed into a set of six irreducible components $I^K_Q$, using 
which we can construct an infinite set of integral equations for their Fourier 
coefficients. Following \citet{hf09} we can decompose 
the Stokes source vector into six irreducible components $S_Q^K$ as 
\begin{equation}
S_{i}(\tau,x,{\bm \Omega}) = \sum_{K=0,2} \ \sum_{Q=-K}^{Q=+K} 
{\mathcal T}^K_Q(i, {\bm \Omega})
\,S^{K}_{Q}(\tau,x,{\bm \Omega}),
\label{si-comp}
\end{equation}
with a similar decomposition for the Stokes vector $I_i$ in terms of 
$I^K_Q$. The ${\mathcal T}^K_Q(i,{\bm \Omega})$ are irreducible spherical tensors 
for polarimetry introduced by \citet{landi84}. 
The irreducible line source vector components are then given by 
\begin{eqnarray}
&&S^{K}_{l,Q}(\tau,x,{\bm \Omega}) =  G^K_Q(\tau)
+ \int_{-\infty}^{+\infty} \md x^\prime 
\oint {\md\Omega^\prime \over {4\pi}}
\nonumber \\ && \times
{1 \over \varphi(x)} \sum_{K^\prime=0,2} \ \sum_{Q^{\prime\prime}=-K^\prime}^
{Q^{\prime\prime}=+K^\prime}
{{\mathcal{R}}}^{KK^\prime}_{QQ^{\prime\prime}}(m,x,x^\prime,\Theta,{\bm B})
\,I^{K^\prime}_{Q^{\prime\prime}}(\tau,x^\prime,{\bm \Omega}^\prime),
\label{skq}
\end{eqnarray}
where $\Theta \ [= \cos^{-1} ({\bm \Omega} . {\bm \Omega^\prime})]$ is the 
scattering angle, $ G^K_Q(\tau)=(\epsilon B_{\nu_0},0,0,0,0,0)^{\rm T}$, and 
${{\mathcal{R}}}^{KK^\prime}_{QQ^{\prime\prime}}$ are the elements of 
angle-dependent 
Hanle redistribution matrix given by 
\begin{eqnarray}
&&{{\mathcal{R}}}^{KK^\prime}_{QQ^{\prime\prime}}(m,x,x^\prime,\Theta,{\bm B})
=\sum_{Q^\prime=-K}^{Q^\prime=+K}\,[{\mathcal N}^K_{QQ^\prime}(m,{\bm B}) 
\,R_{\rm II}(x,x^\prime,\Theta)
\nonumber \\&& +\,{\mathcal N}^{K}_{QQ^\prime}(m,{\bm B})\,
R_{\rm III}(x,x^\prime,\Theta)]\,
\Gamma_{KQ^\prime,K^\prime Q^{\prime\prime}}({\bm \Omega}^\prime).
\label{redismat} 
\end{eqnarray}
In this paper we 
use approximation \rm{II} of \citet{vb97} according to which the redistribution 
matrix is written as a product of magnetic kernel 
${\mathcal N}^K_{QQ^\prime}(m,{\bm B})$ and the angle-dependent redistribution 
functions $R_{\rm II, \rm III}(x,x^\prime,\Theta)$ of \citet{hum62}. 
Here the index $m$ ($=1,\,2,\,3,\,4,\,5$) 
stands for different frequency domains which depend on 
($x$, $x^\prime$, $\Theta$).
The coefficients 
$\Gamma_{KQ^\prime,K^\prime Q^{\prime\prime}}({\bm \Omega}^\prime)$ 
are defined by
\begin{equation}
\Gamma_{KQ^\prime,K^\prime Q^{\prime\prime}}({\bm \Omega}^\prime)
=\sum_{i=0}^{3}\, (-1)^{Q^\prime}\,{\mathcal T}^K_{-Q^\prime}(i,{\bm \Omega}^\prime)
\,{\mathcal T}^{K^\prime}_{Q^{\prime\prime}}(i,{\bm \Omega}^\prime). 
\label{gamma_kqpkqpp}
\end{equation}
The irreducible components $I_Q^K$ and $S_Q^K$ and the coefficients 
$\Gamma_{KQ^\prime,K^\prime Q^{\prime\prime}}$ are complex quantities. For practical 
computations, we prefer working with the real quantities. In order to transfer complex 
quantities into the real space, we follow the procedure given in \citet{hf07}. 
First we define
\begin{eqnarray}
 &&I^{K,\rm x}_Q(\tau,x,{\bm \Omega}) = Re\,\{I^{K}_Q(\tau,x,{\bm \Omega})\},
\nonumber \\&&
I^{K,\rm y}_Q(\tau,x,{\bm \Omega}) = Im\,\{I^{K}_Q(\tau,x,{\bm \Omega})\}.
\label{reim-i}
\end{eqnarray}
Using these real components it can be shown that 
${\bm {\mathcal{S}}}^{\rm r} = (S^0_0, S^2_0, S^{2,\rm x}_1, S^{2,\rm y}_1, 
S^{2,\rm x}_2, S^{2,\rm y}_2)^{\rm T}$ and the corresponding intensity vector 
${\bm {\mathcal{I}}}^{\rm r}$ satisfy the transfer equation given in 
Eq.~(\ref{rte_stokes_component}) with $S_i$ and $I_i$ replaced by 
${\bm {\mathcal{S}}}^{\rm r}$ and ${\bm {\mathcal{I}}}^{\rm r}$ respectively. 
Now using the $\bm{\hat T}$ matrix 
given in Section 5.3 of \citet{hf07} the irreducible line 
source vector in terms of the real quantities can be written as 
\begin{eqnarray}
&&S^{\rm {r},\it {K}}_{l,Q}(\tau,x,{\bm \Omega}) =  G^K_Q(\tau)
+ \int_{-\infty}^{+\infty} \md x^\prime 
\oint {\md\Omega^\prime \over {4\pi}}
\nonumber \\ && \times
{1 \over \varphi(x)} \sum_{K^\prime=0,2} \sum_{Q^{\prime\prime}=0}^
{Q^{\prime\prime}=+K^\prime}
{{\mathcal{R}}}^{\rm{r},\it{KK^\prime}}_{QQ^{\prime\prime}}(m,x,x^\prime,\Theta,{\bm B})
\,I^{\rm {r},\it{K^\prime}}_{Q^{\prime\prime}}(\tau,x^\prime,{\bm \Omega}^\prime).
\label{skq-real}
\end{eqnarray} 
Here 
${{\mathcal{R}}}^{{\rm r}, KK^\prime}_{QQ^{\prime\prime}}$ has the same form as 
Eq.~(\ref{redismat}) with $\mathcal {N}^{K}_{QQ^\prime}$ and 
$\Gamma_{KQ^\prime,K^\prime Q^{\prime\prime}}$ replaced by 
$\mathcal {N}^{\rm{r},\it{K}}_{QQ^\prime}$ and 
$\Gamma^{\rm r}_{KQ^\prime,K^\prime Q^{\prime\prime}}$. 
The $Q$ indices take 
values $[0,+K]$. 
The elements of matrix 
$\Gamma^{\rm r}_{KQ^\prime,K^\prime Q^{\prime\prime}}({\bm \Omega}^\prime)$ are listed in 
Appendix D of \citet{lsaknn11a}. The explicit form of 
$\mathcal {N}^{\rm{r},\it{K}}_{QQ^\prime} (m, \bm{B})$ can be found in 
Appendix A of \citet{lsa11} where they are denoted by ${\bm M}^{(i)} (\bm{B}) $ 
with $i$ playing the role of $m$ in our notation. The formal solution of the transfer equation can 
now be written as
\begin{eqnarray}
 && I^{\rm {r},\it{K}}_Q(\tau,x,{\bm \Omega}) = \int_{\tau}^{+\infty} 
 e^{-(\tau^\prime - \tau) \varphi(x)/\mu} \,S^{\rm {r},\it{K}}_Q(\tau,x,{\bm \Omega}) 
{{\varphi(x)}\over{\mu}} \md \tau^\prime,
\ \ \rm{for} \ \mu > 0,
\nonumber \\&&
I^{{\rm r}, K}_Q(\tau,x,{\bm \Omega}) = -\int_{0}^{\tau} 
 e^{-(\tau^\prime - \tau) \varphi(x)/\mu} \,S^{\rm {r},\it{K}}_Q(\tau,x,{\bm \Omega}) 
{{\varphi(x)}\over{\mu}} \md \tau^\prime,
\ \ \rm{for} \  \mu < 0.
\label{I-formal}
\end{eqnarray}
The irreducible components of the line source vector in Eq.~(\ref{skq-real})
continue to be 
non-axisymmetric, because of the presence of angle-dependent redistribution 
function. 
It is computationally advantageous to express $S^{\rm {r},\it{K}}_{l,Q}$ 
in terms of 
axisymmetric irreducible components. This can be achieved through the 
introduction of Fourier azimuthal expansion of the angle-dependent PRD 
functions. 
 In this paper we use 
approximation II of \citet{vb97}, the expressions of which for the frequency 
domains depend on the scattering angle $\Theta$, and hence on $\bm \Omega$
and $\bm \Omega^\prime$. Therefore, to be consistent, we apply Fourier 
decomposition to the redistribution matrix which contains the 
angle-dependent frequency 
domain information (see \citet{lsaknn12}). This can be done as follows:
\begin{eqnarray}
&&{{\mathcal{R}}}^{{\rm r},KK^\prime}_{QQ^{\prime\prime}}
(m,x,x^\prime,\Theta,{\bm B}) 
= \sum_{k=-\infty}^{k=+\infty}
{\rm e}^{{\rm i}k\chi}\,
\widetilde {{\mathcal{R}}}^{(k) KK^\prime}_{QQ^{\prime\prime}}
(m,x,x^\prime,\theta,{\bm\Omega^\prime},{\bm B}),
\label{azi-r}
\end{eqnarray}
where the Fourier coefficients are given by
\begin{eqnarray}
&&{{\widetilde {\mathcal{R}}}}^{(k) KK^\prime}_{QQ^{\prime\prime}}
(m,x,x^\prime,\theta,{\bm\Omega^\prime},{\bm B}) =  
\int_{0}^{2\pi}{\md \chi\over 2\pi}\,{\rm e}^{{\rm -i}k\chi}\,
{{\mathcal{R}}}^{{\rm r},KK^\prime}_{QQ^{\prime\prime}}(m,x,x^\prime,\Theta,{\bm B}).
\label{four-r}
\end{eqnarray}
The angle-dependent PRD functions $R_{\rm II,\rm III}(x,x^\prime,\Theta)$ 
are periodic 
functions of $\chi$ with a period $2 \pi$ because of which each element of 
the redistribution matrix 
${{\widetilde {\mathcal{R}}}^{(k) KK^\prime}_{QQ^{\prime\prime}}}$ 
is $2 \pi$-periodic.
We remark that in the previous attempts on Fourier decomposition, the expansion 
of angle-dependent functions $R_{\rm II, \rm III}(x,x^\prime,\Theta)$ 
over $(\chi - \chi^\prime)$ was traditionally used 
(see \citet{dh88}, \citet{hf09}, \citet{hf10}). 
We show below that an expansion over $\chi$ of 
the angle-dependent redistribution matrix  
(as done in \citet{lsaknn12}), provides a consistent way of including 
`angle-dependent frequency domains' when performing angle-dependent PRD 
computations. The matrix elements 
${{\widetilde {\mathcal{R}}}^{(k) KK^\prime}_{QQ^{\prime\prime}}}$ 
are studied in detail in Sect.~\ref{azi-rm}. 
Similar azimuthal Fourier expansions for the primary source term 
$G^{K}_{Q}(\tau)$ can be written as 
\begin{equation}
G^K_Q(\tau) = \sum_{k=-\infty}^{k=+\infty}
{\rm e}^{{\rm i}k\chi} \,\,\, \tilde G^{(k)K}_Q(\tau),
\label{azi-G}
\end{equation}
with 
\begin{eqnarray}
&&\tilde G^{(k)K}_Q(\tau) =
\cases{G_0(\tau) & 
if\,\ $k = 0$, \cr
0 & if\,\ $k \neq 0$.}
\label{cond-G}
\end{eqnarray}
Inserting the Fourier azimuthal expansions of 
${{\mathcal{R}}}^{{\rm r},KK^\prime}_{QQ^{\prime\prime}}
(m,x,x^\prime,\Theta,{\bm B})$ 
and  
$G^{K}_{Q}$ in Eq.~(\ref{skq-real}), we obtain an 
expansion for $S^{{\rm r},K}_{l,Q}$ which can be expressed as 
\begin{equation}
 S^{{\rm r},K}_{l,Q}(\tau,x,{\bm \Omega})= \sum_{k=-\infty}^{k=+\infty}
{\rm e}^{{\rm i}k\chi} \,\,\, \tilde S^{(k) K}_{l,Q}(\tau,x,\theta),
\label{azi-S}
\end{equation}
with
\begin{eqnarray}
&&\tilde S^{(k) K}_{l,Q}(\tau,x,{\theta}) = \tilde G^{(k)K}_{Q}(\tau) 
+ \int_{-\infty}^{+\infty}\md x^\prime  \oint 
{\md{\Omega}^\prime \over {4\pi}}
{1\over {\varphi(x)}} 
\nonumber \\ && \times
\sum_{K^\prime=0,2} \sum_{Q^{\prime\prime}=0}^{Q^{\prime\prime}=+K^\prime}
\widetilde {{\mathcal{R}}}^{(k) KK^\prime}_{QQ^{\prime\prime}}
(m,x,x^\prime,\theta,{\bm\Omega^\prime},{\bm B})\,
I^{{\rm r},K^\prime}_{Q^{\prime\prime}}(\tau,x^\prime,\bm\Omega^\prime). 
\label{stildakKQ}
\end{eqnarray}
Substituting from Eq.~(\ref{azi-S}) for $S^{{\rm r},K}_{l,Q}$ in formal 
solution we get
\begin{equation}
 I^{{\rm r},K}_{Q}(\tau,x,{\bm \Omega})= \sum_{k=-\infty}^{k=+\infty}
{\rm e}^{{\rm i}k\chi} \,\,\, \tilde I^{(k)K}_{Q}(\tau,x,\theta),
\label{azi-I}
\end{equation} 
where
\begin{eqnarray}
 && \tilde I^{(k)K}_Q(\tau,x, \theta) = \int_{\tau}^{+\infty} 
 e^{-(\tau^\prime - \tau) \varphi(x)/\mu} \,
\tilde S^{(k)K}_Q(\tau,x,{\theta})\, 
{{\varphi(x)}\over{\mu}} \,\md \tau^\prime,
\ \ \rm{for} \ \mu > 0,
\nonumber \\&&
\tilde I^{(k)K}_Q(\tau,x,{\theta}) = -\int_{0}^{\tau} 
 e^{-(\tau^\prime - \tau) \varphi(x)/\mu} \,
\tilde S^{(k)K}_Q(\tau,x,\theta) \,
{{\varphi(x)}\over{\mu}}\, \md \tau^\prime,
\ \ \rm{for} \ \mu < 0.
\label{I-tilda}
\end{eqnarray}
Thus from Eqs.~(\ref{stildakKQ}) and (\ref{azi-I}) we get an expression 
for azimuthal Fourier source vector components as
\begin{eqnarray}
&&\tilde S^{(k)K}_{l,Q}(\tau,x,{\theta}) = \tilde G^{(k)K}_{Q}(\tau) 
+ \int_{-\infty}^{+\infty}\md x^\prime  \oint 
{\md{\Omega}^\prime \over {4\pi}} 
{1\over {\varphi(x)}} 
\nonumber \\ && \times
\sum_{K^\prime=0,2} \sum_{Q^{\prime\prime}=0}^
{Q^{\prime\prime}=+K^\prime}
\widetilde {{\mathcal{R}}}^{(k) KK^\prime}_{QQ^{\prime\prime}}
(m,x,x^\prime,\theta,{\bm\Omega^\prime},{\bm B})
\,\sum_{k^\prime =-\infty}^{k^\prime=+\infty} 
{\rm e}^{{\rm i}k^\prime \chi^\prime}
\,\tilde I^{(k^\prime)K^\prime}_{Q^{\prime\prime}}
(\tau,x^\prime,\theta^\prime). 
\label{skKQ-final}
\end{eqnarray}
Notice that the Fourier indices 
$k$ and $k^\prime$ are not 
coupled to $Q$ and ${Q^{\prime\prime}}$ unlike in the case of 
decomposition over 
$(\chi - \chi^\prime)$ (see \citet{hf09}, NS11).\\ 
\indent The advantage 
of working in real irreducible basis is that we can reduce the computational time by 
restricting the values of azimuthal Fourier index $k$ to positive space using 
conjugate symmetry relations as shown below. This 
simplification is analytically complicated in complex basis.
From Eq.~(\ref{four-r}) we can see that the components 
$\widetilde {{\mathcal{R}}}^{(k)KK^\prime}_{QQ^{\prime\prime}}$ 
satisfy the symmetry relation 
\begin{equation}
\widetilde {{\mathcal{R}}}^{(k)KK^\prime}_{QQ^{\prime\prime}}
=\bigg[\widetilde {{\mathcal{R}}}^{(-k)KK^\prime}_{QQ^{\prime\prime}}\bigg]^*.
\label{}
\end{equation}
Using the above relation in Eq.~(\ref{azi-r}) we get
\begin{eqnarray}
&&{{\mathcal{R}}}^{{\rm r},KK^\prime}_{QQ^{\prime\prime}}
(m,x,x^\prime,\Theta,{\bm B})= Re\, \bigg[ \,\sum_{k=0}^{k=+\infty}
(2-\delta_{k0})\,
{\rm e}^{{\rm i}k\chi}\,
\widetilde {{\mathcal{R}}}^{(k)KK^\prime}_{QQ^{\prime\prime}}
(m,x,x^\prime,\theta,{\bm\Omega^\prime},{\bm B}) \bigg].
\label{r-positive}
\end{eqnarray}
Notice that the Fourier series constitutes only the terms with 
$k\geqslant0$ which is useful in practical computations. With this 
simplification and following as in Eqs.~(\ref{azi-S})-(\ref{skKQ-final}) 
we get
\begin{equation}
 S^{{\rm r},K}_{l,Q}(\tau,x,{\bm \Omega})= Re\,\left[ \sum_{k=0}^{k=+\infty}
(2-\delta_{k0})\,
{\rm e}^{{\rm i}k\chi} \, \tilde S^{(k)K}_{l,Q}(\tau,x,\theta)\right],
\label{S-positive}
\end{equation}
and
\begin{equation}
 I^{{\rm r},K}_{l,Q}(\tau,x,{\bm \Omega})= Re\,\left[ \sum_{k=0}^{k=+\infty}
(2-\delta_{k0})\,
{\rm e}^{{\rm i}k\chi} \, \tilde I^{(k)K}_{l,Q}(\tau,x,\theta)\right].
\label{I-positive}
\end{equation}
The $\tilde S^{(k)K}_{l,Q}(\tau,x,{\bm \Omega})$ now takes the form 
\begin{eqnarray}
&&\tilde S^{(k)K}_{l,Q}(\tau,x,{\theta}) = \tilde G^{(k)K}_{Q}(\tau) 
+ \int_{-\infty}^{+\infty}\md x^\prime  \oint 
{\md{\Omega}^\prime \over {4\pi}}
{1\over {\varphi(x)}} \sum_{K^\prime=0,2} \sum_{Q^{\prime\prime}=0}^
{Q^{\prime\prime}=+K^\prime}
\widetilde {{\mathcal{R}}}^{(k)KK^\prime}_{QQ^{\prime\prime}}
(m,x,x^\prime,\theta,{\bm\Omega^\prime},{\bm B})
\nonumber \\ && \times
Re \left[ \sum_{k^\prime =0}^{k^\prime=+\infty} (2-\delta_{k^\prime0}) 
{\rm e}^{{\rm i}k^\prime \chi^\prime}
\,\tilde I^{(k^\prime)K^\prime}_{Q^{\prime\prime}}
(\tau,x^\prime,\theta^\prime) \right]. 
\label{skKQ-positive}
\end{eqnarray}

\section{Azimuthal Fourier components of the redistribution matrix elements}
\label{azi-rm}
\begin{figure*}
\centering
\includegraphics[height=6.2cm,width=4.2cm]{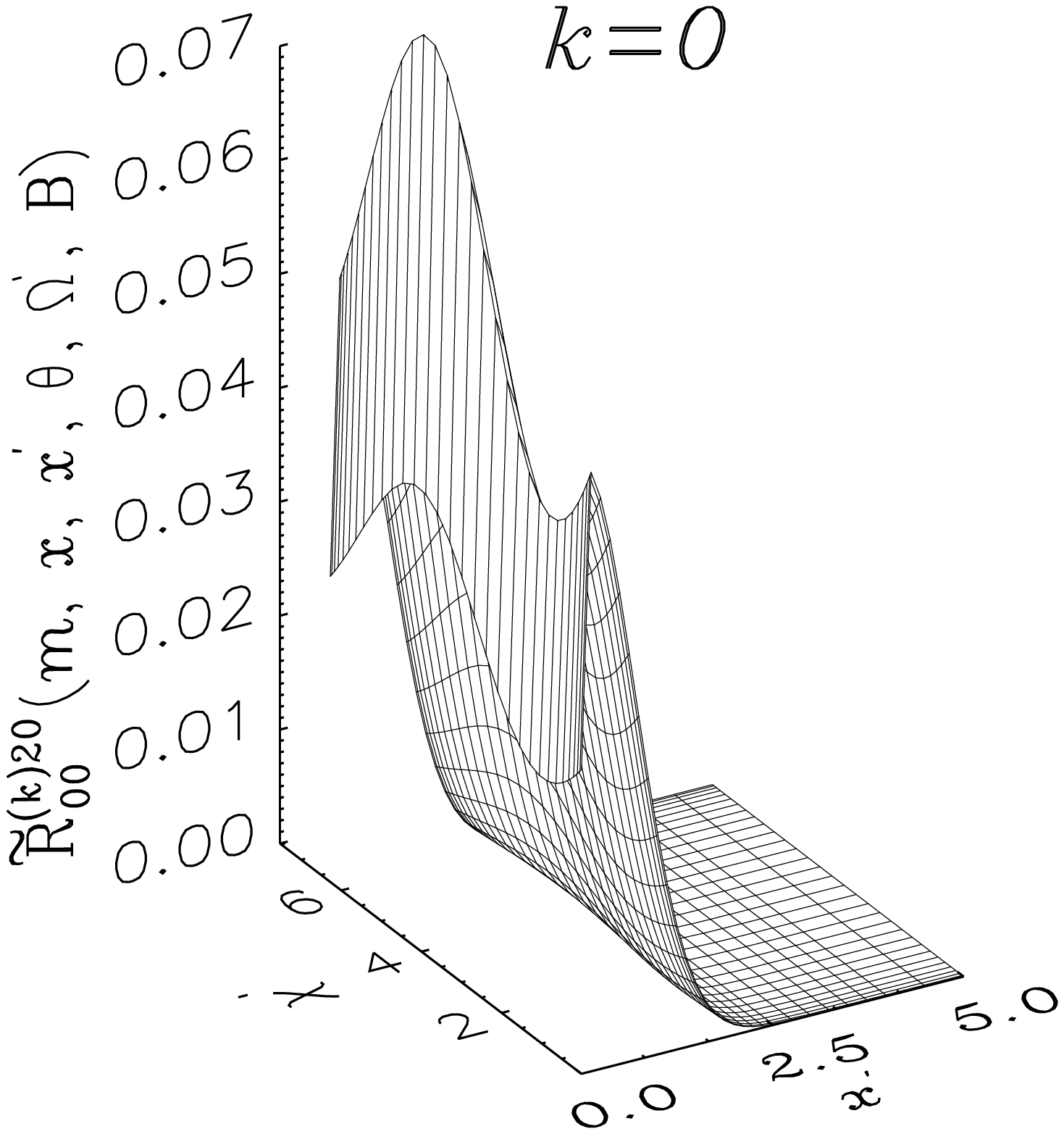}
\includegraphics[height=6.2cm,width=4.2cm]{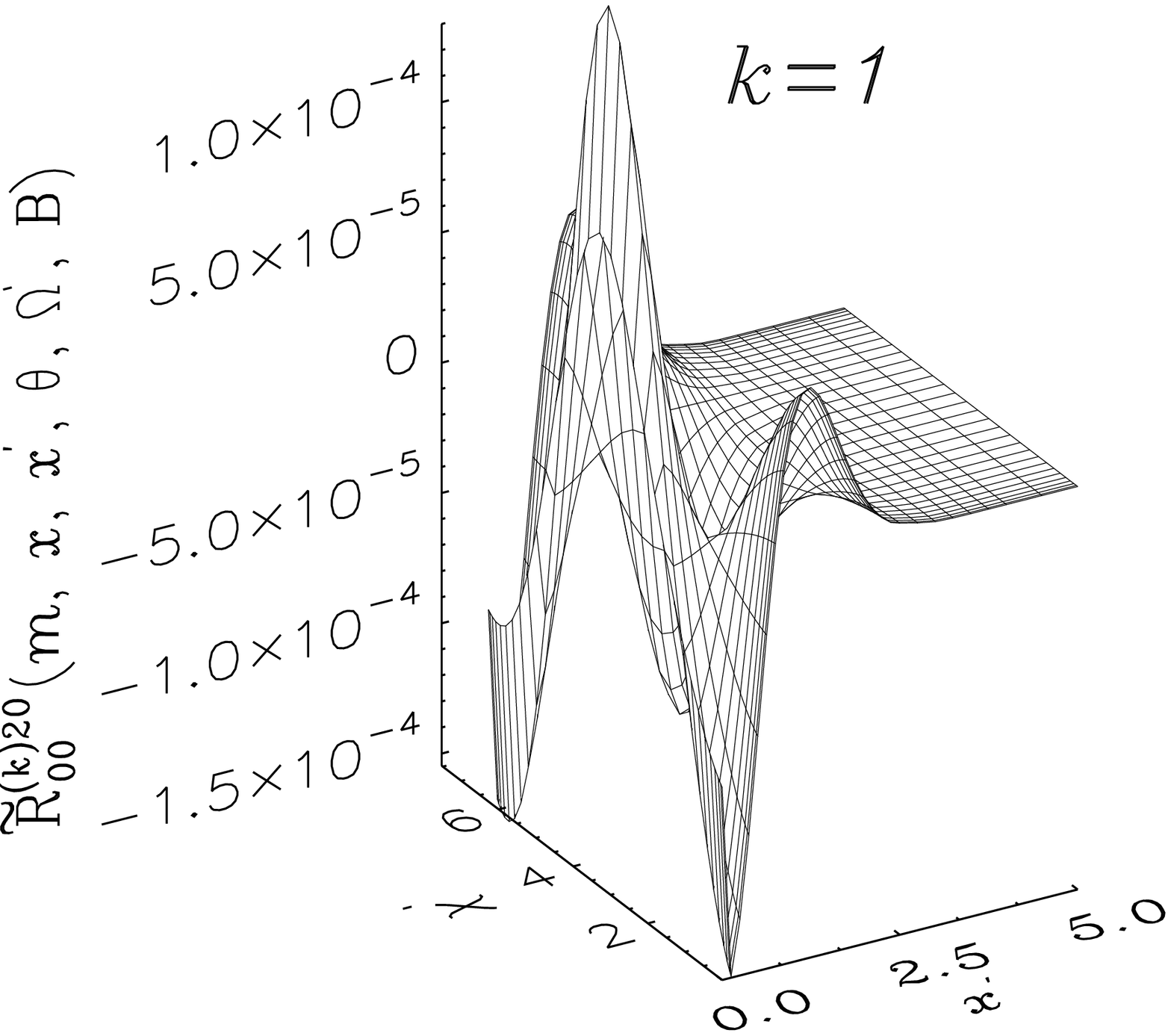}
\includegraphics[height=6.2cm,width=4.2cm]{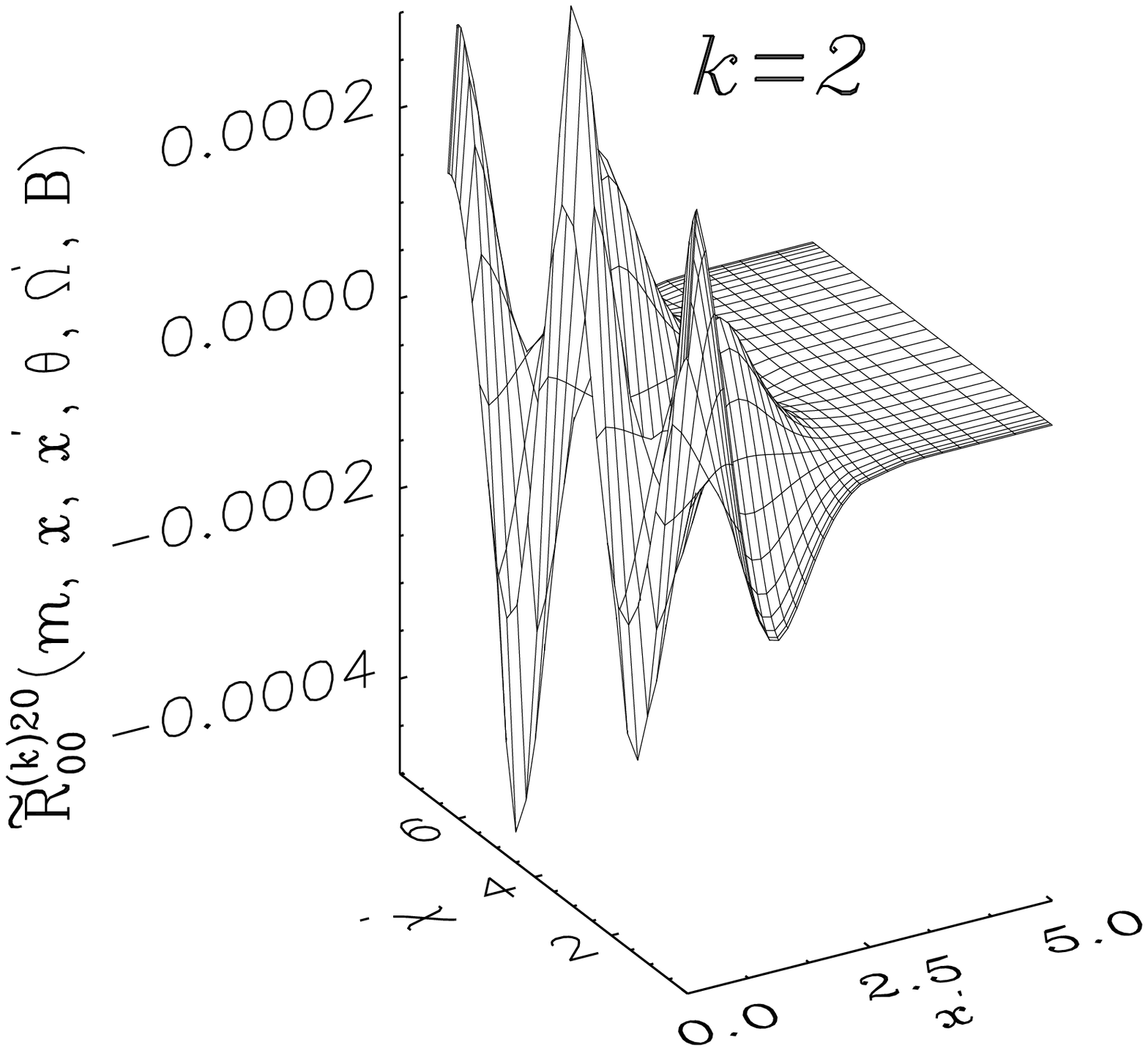}
\includegraphics[height=6.2cm,width=4.2cm]{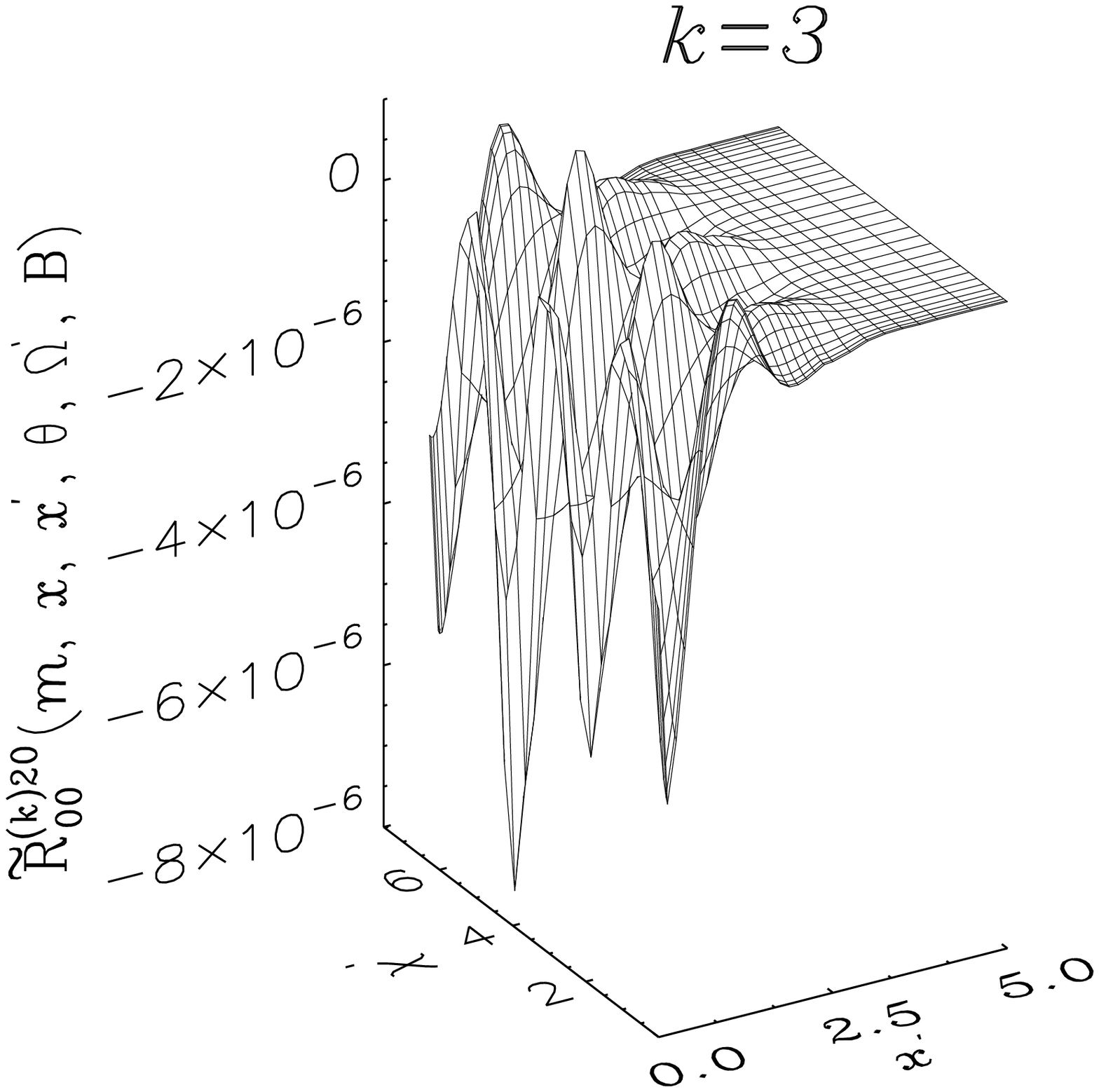}
\includegraphics[height=6.2cm,width=4.2cm]{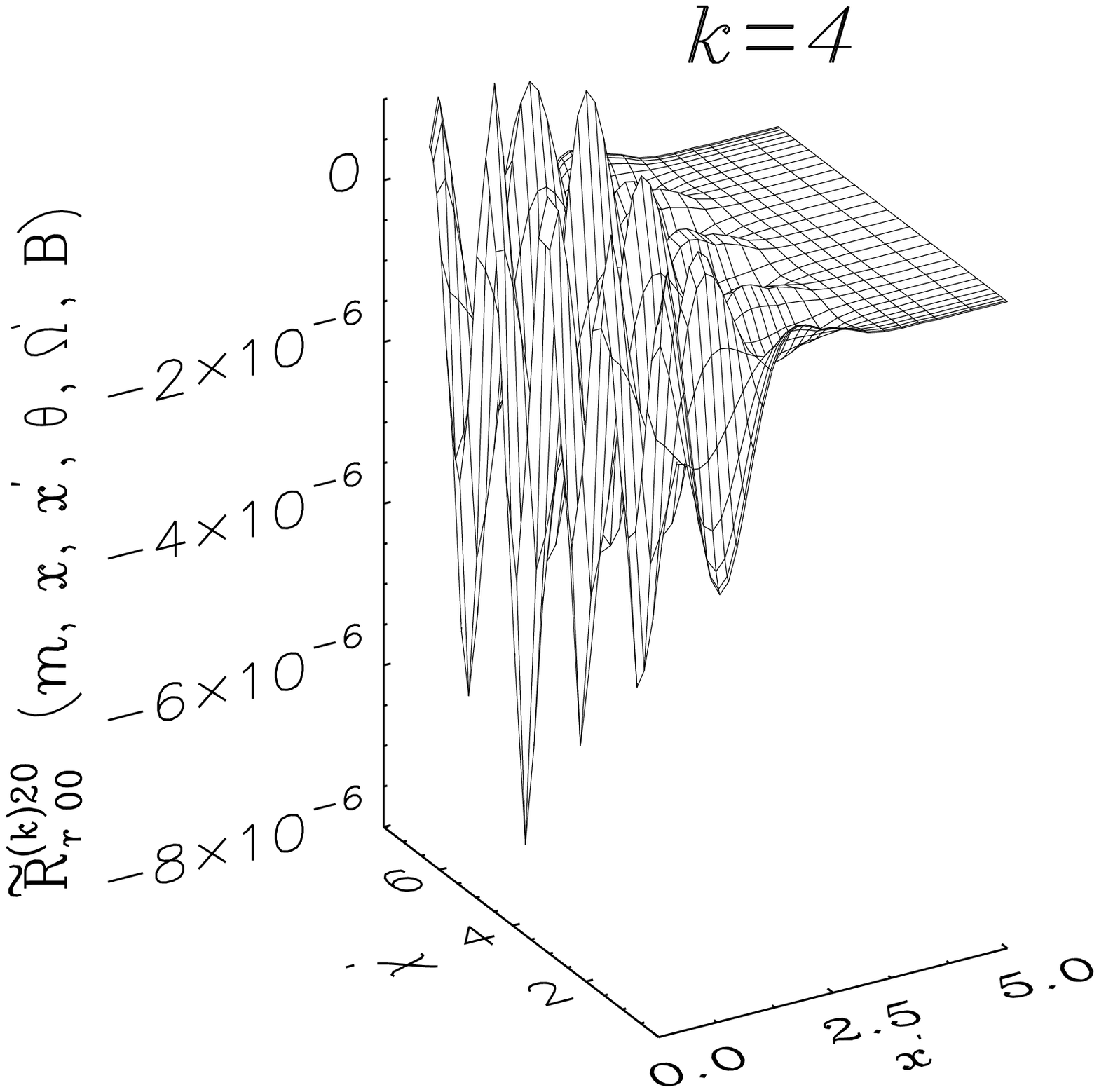}
\includegraphics[height=6.2cm,width=4.2cm]{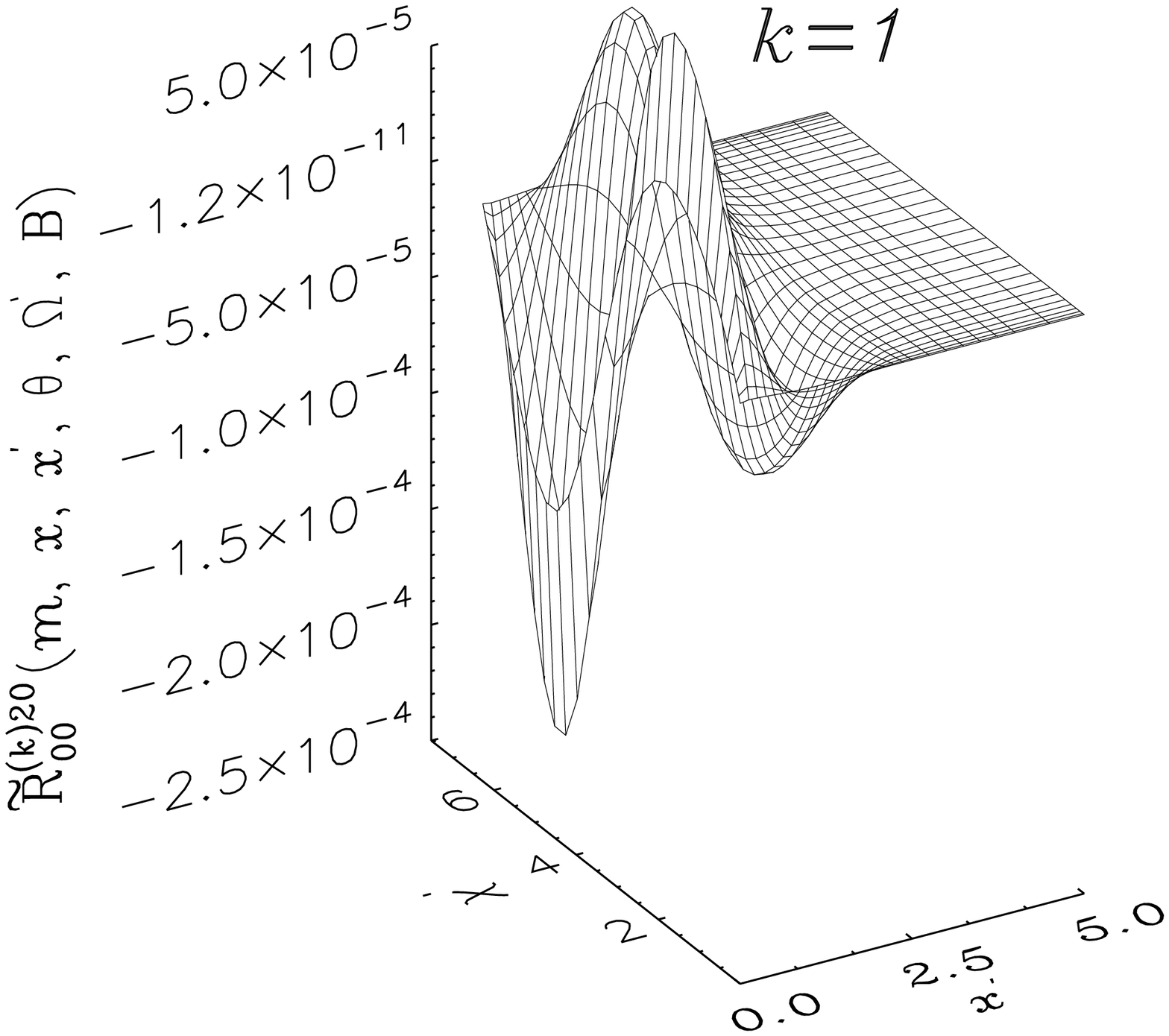}
\includegraphics[height=6.2cm,width=4.2cm]{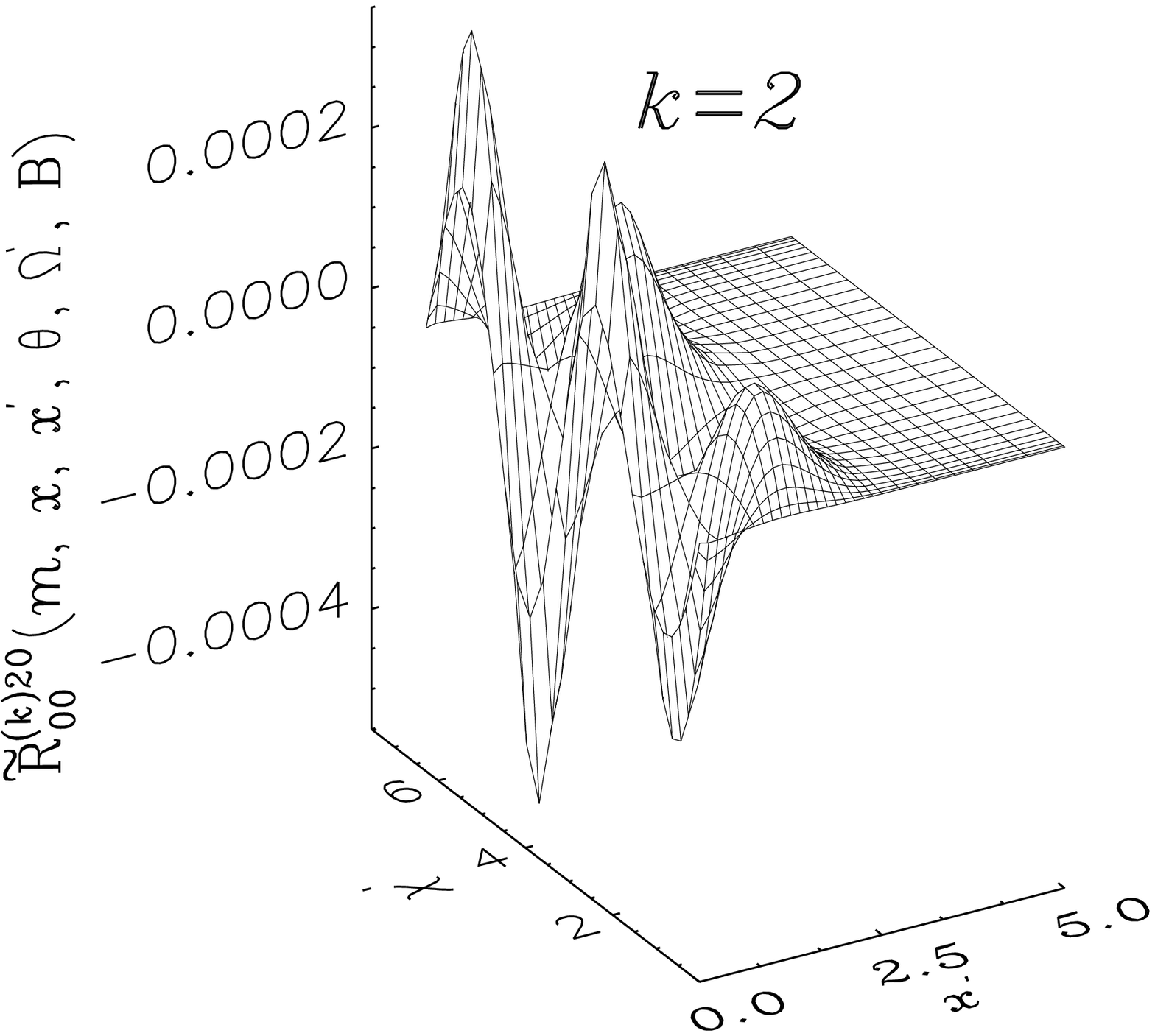}
\includegraphics[height=6.2cm,width=4.2cm]{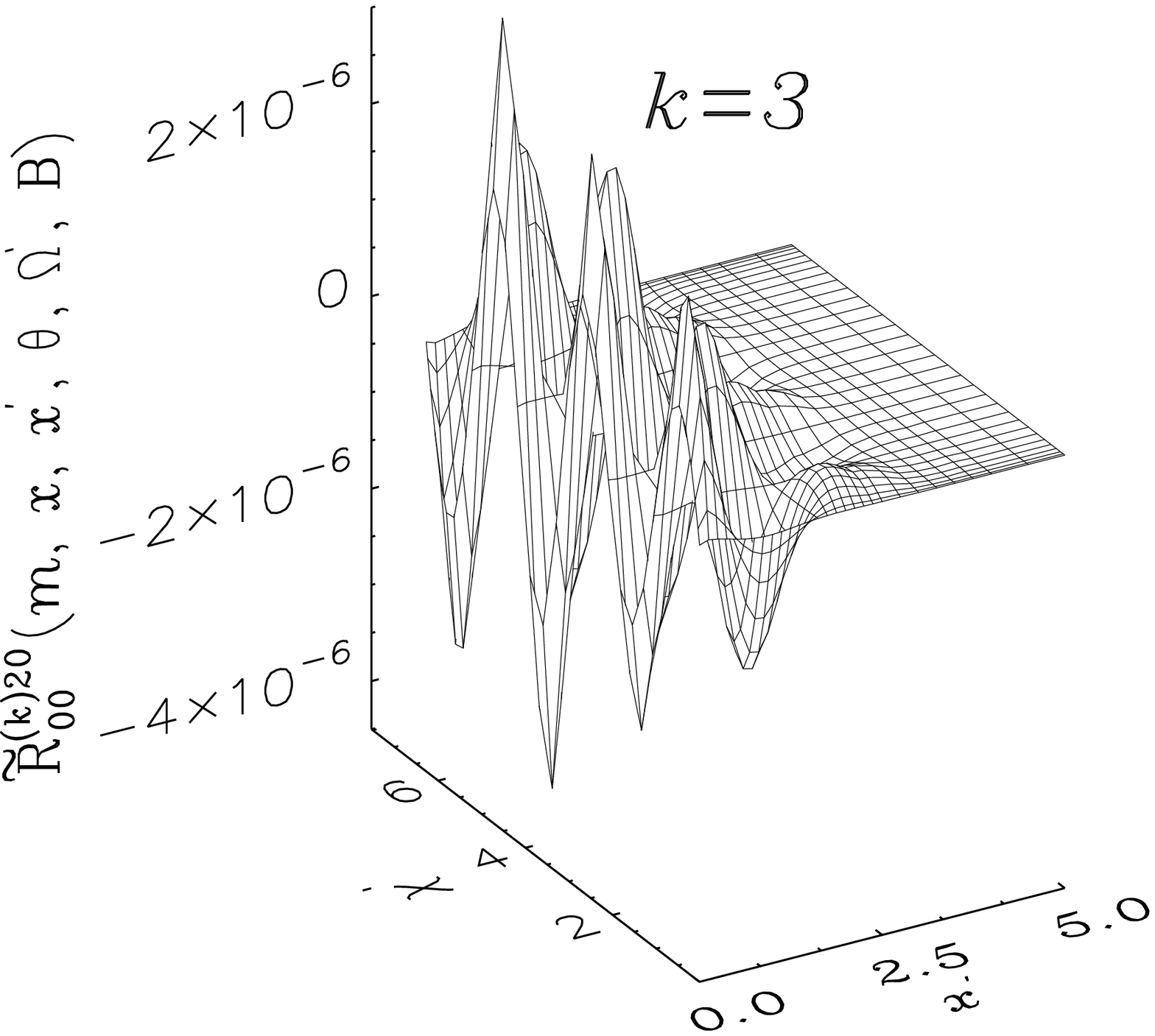}
\includegraphics[height=6.2cm,width=4.2cm]{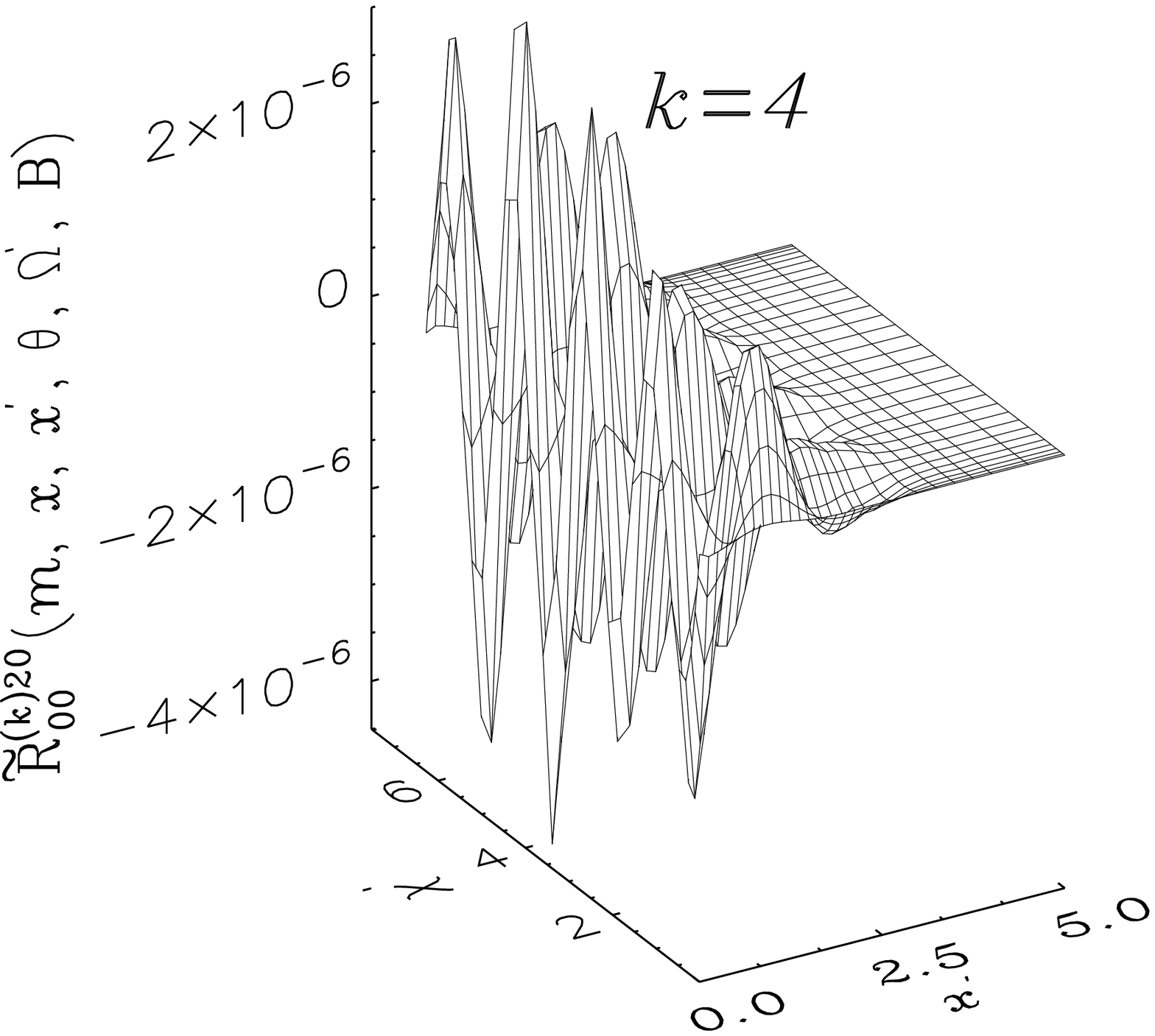}
\caption{Fourier azimuthal components of the type \rm{II} 
redistribution matrix elements are shown as a function of $x^\prime$ and 
$\chi^\prime$ for $x=0$, $\theta=\pi/2$, $\theta^\prime=\pi$ and 
($\Gamma$, $\theta_B$, $\chi_B$) = (1, 30{\textdegree}, 0{\textdegree}). 
The first five panels from left to right correspond to the real part of 
$\widetilde {{\mathcal{R}}}^{(k)20}_{00}$ 
and the remaining panels refer to the imaginary part of 
$\widetilde {{\mathcal{R}}}^{(k)20}_{00}$.}
\label{azi-rmelements}
\end{figure*}
In this section we present the azimuth angle 
dependence of the Fourier decomposed matrix elements of the redistribution 
matrix, namely $\widetilde {{\mathcal{R}}}^{(k)KK^\prime}_{QQ^{\prime\prime}}
(m,x,x^\prime,\theta,{\bm\Omega^\prime},{\bm B})$. From Eq.~(\ref{r-positive}), 
it is clear that the value of $k$ extends from 
 $0$ to $+\infty$. For numerical evaluation it is necessary to truncate 
this infinite series. Our studies show that 
the series can be truncated at $k=4$. We compute Fourier components 
$\widetilde {{\mathcal{R}}}^{(k)KK^\prime}_{QQ^{\prime\prime}}$ numerically. This
we do by numerical integration of 
${{\mathcal{R}}}^{KK^\prime}_{QQ^{\prime\prime}}$ over the azimuth angle $\chi$ 
using a Gauss-Legendre quadrature with 32 grid points between $[0,2\pi]$.\\
\indent Fig.~\ref{azi-rmelements} shows the 
$\widetilde {{\mathcal{R}}}^{(k)20}_{00}$ element of the 
redistribution matrix as a function of $x^\prime$ and $\chi^\prime$ for a 
given set of $x$, $\theta$, $\theta^\prime$ and the magnetic field parameters. 
The main feature is that the $k=0$ component is the dominant term. Even though 
$k\neq0$ terms depend sensitively on $\chi^\prime$, their magnitudes are several 
orders smaller than the $k=0$ component. For this reason, we can 
truncate the Fourier azimuthal expansion of the redistribution matrix elements 
to the fifth term itself without causing significant errors. In fact for practical 
computation one can truncate the series at $k=2$ itself. This would help in rapid 
computation of the angle-dependent PRD problems in practical applications. 
However, in the theoretical studies presented in the paper we use $k=4$. From 
Fig.~\ref{azi-rmelements} it follows that the higher order components show several 
harmonics as $\chi^\prime$ varies from $0$ to $2\pi$. This behavior is confined 
to $x^\prime \lesssim 3$ when $x = 0$. For larger values of $x^\prime$, the 
components approach the value zero. We have verified that the above 
conclusions remain valid for arbitrary choice of $x$, $\theta$, $\theta^\prime$ 
and the magnetic field parameters and for other combinations of $K$, $K^\prime$, 
$Q$, $Q^{\prime\prime}$.

\section{Scattering expansion method for Hanle effect with angle-dependent 
PRD}
\label{SEM}
In Sect.~\ref{azi-rm} we showed that the azimuthal Fourier expansion of the 
redistribution matrix (see Eq.~(\ref{r-positive})) can be truncated to the 
fifth term. Thus, for $k=0,\,1,\,2,\,3,\,4$, we obtain a finite set of 54 coupled 
integral equations. The dimensionality of the problem increases to 54 
complex quantities  
from 54 real quantities when we work in full space 
(i.e., $-4 \leqslant k \leqslant +4$). Thus working in positive half space is 
computationally advantageous. 
In this section we present an iterative method to solve this set of coupled 
equations. This method is based on Neumann series 
expansion of the components of the source vector contributing to the 
polarization. \citet{sametal11} applied this method to solve the transfer problem 
with angle-dependent PRD in non-magnetic case and named it  
as SEM. These authors also show the efficiency of SEM over the core-wing-based 
polarized approximate lambda iteration (ALI) method. NS11 employed SEM to solve Hanle transfer problem with 
angle-dependent PRD. The results obtained by them showed slight inconsistency as 
compared to the results obtained from perturbation method (\citet{knnetal02}). 
This might be due to the use of angle-averaged frequency domains 
(approximation \rm{III}) to solve Hanle transfer problem with angle-dependent 
PRD functions. Consistency in the results can be obtained by 
actually using the angle-dependent 
frequency domains (approximation \rm{II}). \\
\indent In SEM, first neglecting polarization we calculate Stokes $I$. We assume 
that Stokes $I$ is cylindrically symmetric and is given by the component 
$\tilde{I}^{(0)0}_0$ itself to an excellent approximation. This 
approximation yields $k^\prime = K^\prime = Q^{\prime\prime} = 0$ in 
Eq.~(\ref{skKQ-positive}). The resulting component is the solution of a non-LTE 
unpolarized radiative transfer equation with the line source function given by 
\begin{eqnarray}
&&\tilde S^{(0)0}_{l,0}(\tau,x,\theta)=\epsilon B_{\nu_0} +
\int_{-\infty}^{+\infty}\md x^\prime \oint {\md{{\Omega}^\prime}
 \over {4 \pi} }\,
{1 \over {\varphi(x)}}\,
\widetilde {{\mathcal{R}}}^{(0)00}_{00}
(m,x,x^\prime,\theta,{\bm\Omega^\prime},{\bm B}) 
\,\tilde {I}^{(0)0}_0(\tau,x^\prime,\theta^\prime).
\label{s00_approx}
\end{eqnarray}
Eq.~(\ref{s00_approx}) can be solved using a scalar ALI method based on a 
core-wing approach.
Keeping only the contribution of  $\tilde{I}^{(0)0}_0$ on the 
RHS of Eq.~(\ref{skKQ-positive}) to the $K=2$ coefficients and 
$k=0,\,1,\,2,\,3,\,4$, each component $\tilde S^{(k)2}_{l,Q}$ can be 
written as 
\begin{eqnarray}
&&\left[\tilde S^{(k)2}_{l,Q}(\tau,x,\theta)\right]^{(1)} 
\simeq 
\int_{-\infty}^{+\infty}\md x^\prime \oint {\md{{\Omega}^\prime}
 \over {4 \pi} }\,
{1\over {\varphi(x)}} \,
\widetilde {{\mathcal{R}}}^{(k)20}_{Q0}
(m,x,x^\prime,\theta,{\bm\Omega^\prime},{\bm B})
\,\tilde {I}^{(0)0}_0(\tau,x^\prime,\theta^\prime).
\label{skKlQ_approx}
\end{eqnarray}
The superscript 1 stands for the 
single scattering approximation to the polarized component of the 
source vector. The corresponding radiation field 
$\left[\tilde{I}^{(k)2}_Q\right]^{(1)}$ for $k=0,\,1,\,2,\,3,\,4$ 
is calculated with a formal solver and it serves as a starting solution
for calculating the higher-order terms. 
The higher-order terms can be 
obtained by substituting for 
$\tilde I^{(k^\prime)2}_{Q^{\prime\prime}}$ 
appearing in the RHS of Eq.~(\ref{skKQ-positive}), from 
$\left[\tilde{I}^{(k)2}_{Q}\right]^{(1)}$.
We see that indices $k, k^\prime, Q$ and $Q^\prime$ are now 
decoupled whereas they were coupled in the case of Fourier decomposition over 
$(\chi - \chi^{\prime})$. Correspondingly the number of non-zero 
$\left[\tilde{I}^{(k)2}_{Q}\right]^{(1)}$ has increased from 
25 to 54 when changing from Fourier expansion over 
$(\chi - \chi^{\prime})$ to that over $\chi$. As a result the 
dimensionality of the problem has increased in the single-scattered solution computation.\\
\indent In the computation of higher order scattering terms, apart from keeping 
the coupling of $(K=2,Q)$ components with other polarization components 
$(K^\prime=2,Q^{\prime\prime})$, we also keep the coupling of $k$ 
with all other $k^\prime$ terms. We recall that in NS11 coupling of $k$ 
with $k^\prime = 0$ terms were only retained. Thus  
$\tilde S^{(k)2}_{l,Q}$ at order {\it n} are now given by 
\begin{eqnarray}
&&\left[\tilde S^{(k)2}_{l,Q}(\tau,x,\theta)\right]^{(n)} 
\simeq \left[\tilde S^{(k)2}_{l,Q}(\tau,x,\theta)\right]^{(1)} 
+
\int_{-\infty}^{+\infty}\md x^\prime \oint {\md {\Omega}^\prime
\over {4\pi}}\,
{1\over {\varphi(x)}}
\nonumber \\ && \times
\sum_{Q^{\prime\prime}=0}^{Q^{\prime\prime}=2}
\widetilde {{\mathcal{R}}}^{(k)22}_{QQ^{\prime\prime}}
(m,x,x^\prime,\theta,{\bm\Omega^\prime},{\bm B})
\,Re \,\bigg\{ \sum_{k^\prime = 0}^{k^\prime = +\infty} (2-\delta_{k^\prime0})\,
{\rm e}^{{\rm i}k^\prime \chi^\prime}\,
\left[\,\tilde I^{(k^\prime)2}_{Q^{\prime\prime}}(\tau,x^\prime,\theta^\prime)
\,\right]^{(n-1)}\bigg\}. 
\label{skKlQ_sem}
\end{eqnarray}
From Fig.~\ref{azi-rmelements} it can be seen that $k=0$ component of the 
redistribution matrix elements dominate over the higher order terms ($k\neq0$). 
For this reason, despite a strong dependence of 
$\widetilde {{\mathcal{R}}}^{(k)KK^\prime}_{QQ^{\prime\prime}}$ on $\chi^\prime$, 
it is sufficient in the summation over $k^\prime$ to retain the leading term 
(namely $k^\prime = 0$) in practical computations. The inclusion of higher order 
terms $k^\prime > 0$ do not affect the solutions significantly. Using only 
$k^\prime = 0$ term also saves great amount of computing effort. 
\begin{figure*}
\centering
\includegraphics[height=12.0cm,width=7.5cm]{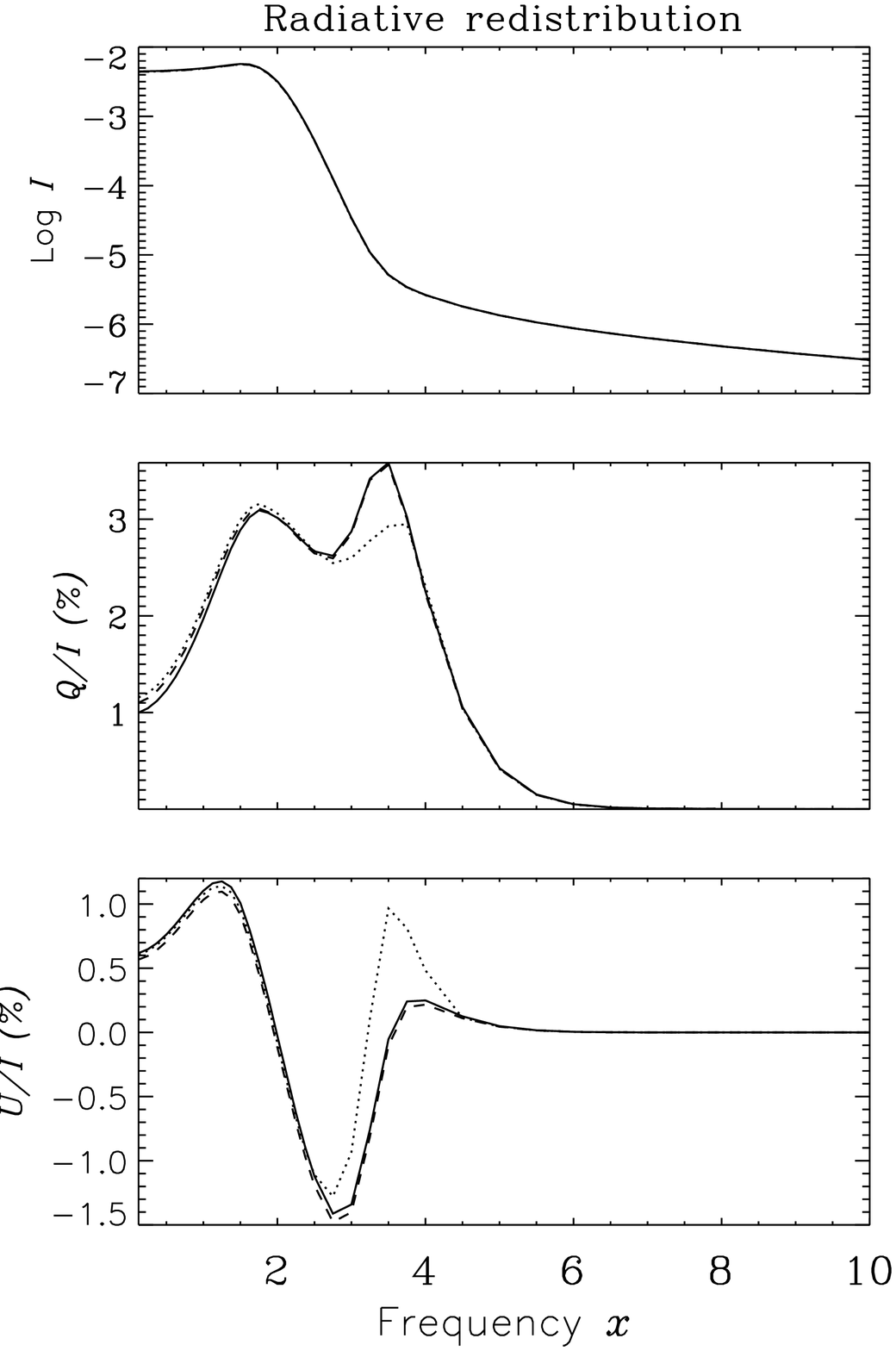}
\includegraphics[height=12.0cm,width=7.5cm]{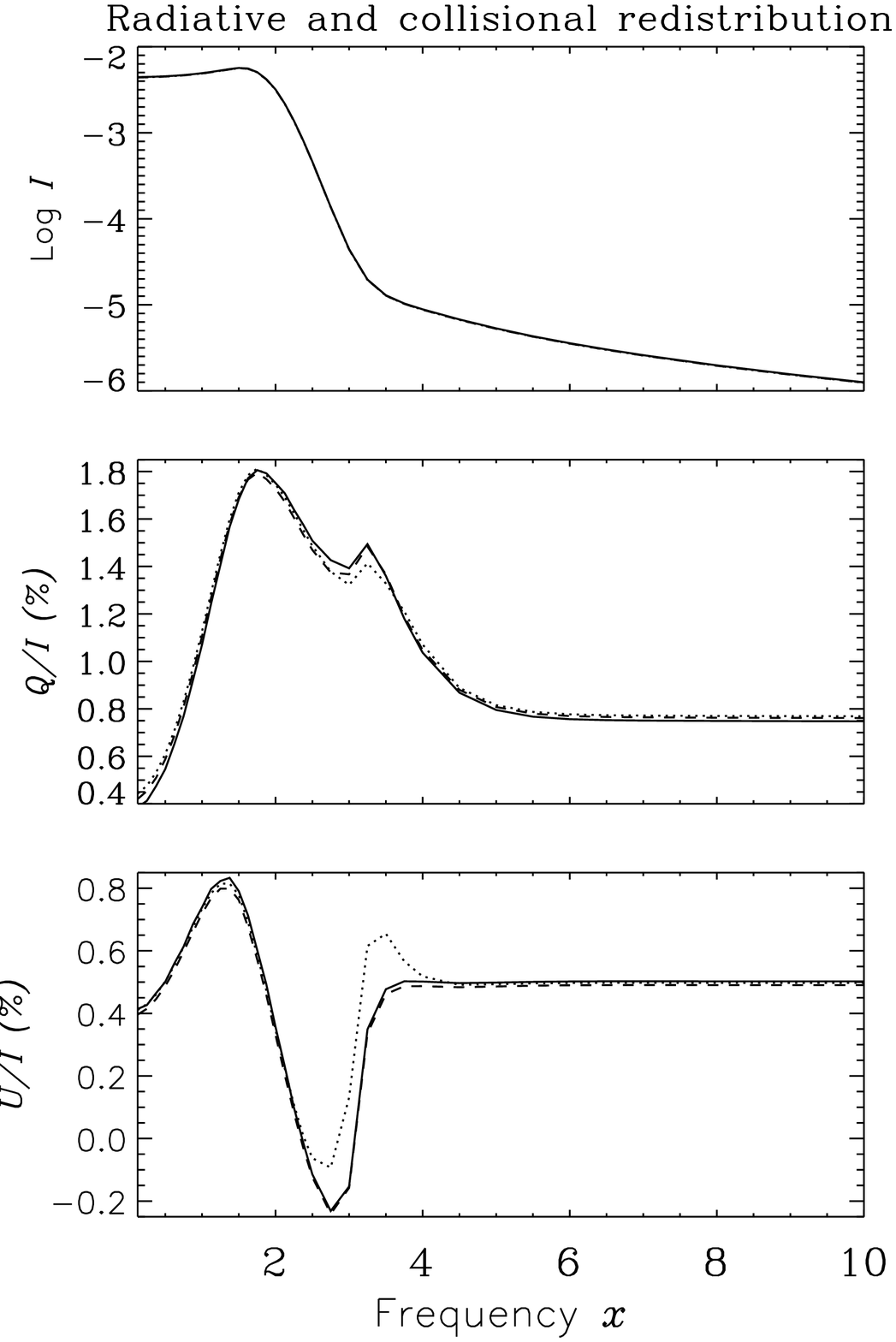}
\caption{Stokes profiles at $\mu = 0.112$ computed using three numerical methods, 
namely the 
perturbation method (solid line), the NS11 approach (dotted line), and 
the self consistent approach used in this paper (dashed line). Model parameters are 
($T$, $a$, $\epsilon$, $r$, $B_{\nu_0}$) = 
($10$, $10^{-3}$, $10^{-3}$, 0, 1). The magnetic field parameters are taken as 
($\Gamma$, $\theta_B$, $\chi_B$) = (1, 30{\textdegree}, 0{\textdegree}). 
The left panel shows the results computed 
using pure $R_{\rm II}$ function 
and the right panel is for a combination of $R_{\rm II}$ and $R_{\rm III}$ 
functions with $\Gamma_{E}/\Gamma_{R}$ = 1.}
\label{comp-t10}
\end{figure*}
\section{Results and discussions}
\label{results}
We compare the Stokes parameter $I$ and the ratios $Q/I$ and $U/I$ computed 
from our present approach with the perturbation method of \citet{knnetal02} 
and NS11 approach. 
The perturbation method treats linear polarization as a perturbation to the 
scalar intensity, and computes the polarization in a two step process, wherein 
an accurately computed Stokes $I$ is used as an input in evaluation of polarized 
scattering integral. In successive perturbations, the Stokes $Q$ and $U$ are 
computed more and more accurately until convergence is reached. NS11 
approach is discussed in Sections ~\ref{sec_intro} and ~\ref{comp}. 
We consider 
self-emitting plane-parallel, isothermal atmospheres with no incident 
radiation at the boundaries. The slab models are characterized by 
($T$, $a$, $\epsilon$, $r$, $B_{\nu_0}$, $\Gamma_{E}/\Gamma_{R}$ ), 
where $T$ is the optical thickness of the slab, and $\Gamma_{E}$ is the 
elastic collision rate. The depolarizing collision rate $D^{(2)}$ is set 
to $\Gamma_{E}/2$. 
The plank function $B_{\nu_0}$ is taken as unity at the line center. 
The polarizability factor $W_2$ is taken as unity (i.e, we consider a 
$J=0\rightarrow1\rightarrow0$ scattering transition with $J$ the total 
angular momentum quantum number). The vector magnetic field in the Hanle scattering problem is defined through 
the field strength parameter $\Gamma = g\omega_L / \Gamma_R$, with $g$ the
Land\'e factor of the upper level and $\omega_L$ the  
Larmor frequency; the field inclination $(\theta_B, \chi_B)$ defined 
with respect to the atmospheric normal. 
For angle and frequency discretization we have used quadratures of the 
same order as those used by NS11. 
Therefore we do not elaborate here on the computational aspects.
\subsection{A comparison with previous approaches to solve the 
angle-dependent Hanle transfer problem}
\label{comp}
\begin{figure*}
\centering
\includegraphics[height=12.0cm,width=8.5cm]{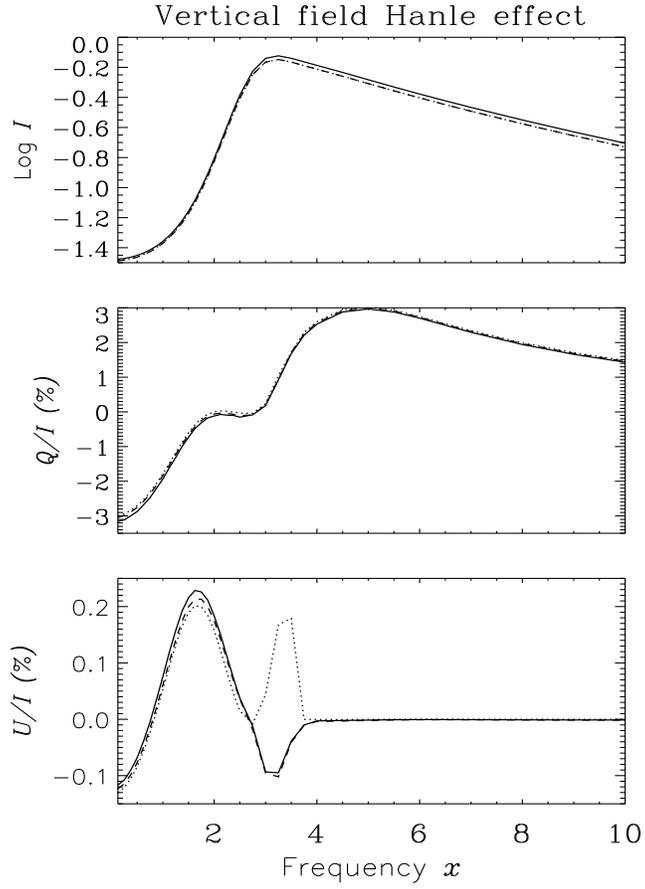}
\caption{A comparison of emergent Stokes profiles computed by 
three numerical methods discussed in the text. The profiles 
are presented for $\mu = 0.112$ and for a vertical 
magnetic field ($\Gamma$, $\theta_B$, $\chi_B$) = 
(1, 0{\textdegree}, 0{\textdegree}). 
Different line types represent same cases as in 
Fig.~\ref{comp-t10}. 
The slab model parameters are 
($T$, $a$, $\epsilon$, $r$, $B_{\nu_0}$, $\Gamma_{E}/\Gamma_{R}$) = 
($2\times10^4$, $10^{-3}$, $10^{-3}$, 0, 1, 1). }
\label{comp-vh}
\end{figure*}
In this section we present the Stokes profiles of the lines formed in a 
magnetized slab scattering according to Hanle PRD matrix 
formulated by \citet{vb97}. 
In the so called approximation \rm{II} and \rm{III} of 
\citet{vb97}, the switch over from the Hanle 
phase matrices (in the core) to the Rayleigh phase 
matrix (in the wings) is achieved through the use of angle-dependent 
and angle-averaged frequency domains respectively. It is shown by \citet{vb97} that the use of 
frequency domains simplifies the numerical evaluation of the 
redistribution matrices.\\
\indent  It is natural that in AD radiative transfer 
computations involving AD functions, one should use Approximation \rm{II} 
involving AD functions. 
However, as already discussed in Sects.~\ref{sec_intro} and \ref{SEM}, 
in NS11 AA domains were 
used in computing the AD redistribution matrix. Although such an approach 
is inconsistent, it provides a rapid means of solving the Hanle AD PRD problems. 
In the present paper we test their approach by actually using AD frequency domains while 
computing AD redistribution matrix (which is fully consistent). 
In Fig.~\ref{comp-t10}, 
we show a comparison of results obtained by the NS11 and the present approach, 
along with those obtained from the simple perturbation method 
of \citet{knnetal02}, 
which is also consistent, like the present 
approach. The left panels in Fig.~\ref{comp-t10} show the 
Stokes profiles computed using 
pure $R_{\rm{II}}$ function. 
One can clearly see that the NS11 approach differs from the present approach 
particularly in the frequency range $3 \lesssim x \lesssim 5$. The present approach 
and perturbation method give same results. The impact of the approximation 
used in the NS11 is more severe on the Stokes $U$ parameter. 
The right panels in Fig.~\ref{comp-t10} show the  
results computed using the same model as in the left panels, 
but for the introduction 
of elastic collisions (a combination of $R_{\rm II}$ and $R_{\rm III}$). 
One can clearly see that the 
differences between NS11 and the present approach still exist, although the 
collisions decrease these differences. The present approach, 
unlike the approximate treatment followed in NS11, thus provides a self-consistent approach 
to compute the redistribution matrix 
(i.e., the use of AD domains to compute AD redistribution matrices), 
at the same time requiring 
manageable computing resources . This has practical implications in realistic 
modeling of the observed Stokes profiles.\\
\begin{table}[ht]
 \begin{centering}
\caption{Comparison of CPU time taken by different methods for radiative 
transfer computations. The model parameters used for the computation are 
($T$, $a$, $\epsilon$, $r$, $B_{\nu_0}$, $\Gamma_{E}/\Gamma_{R}$) = 
($2\times10^6$, $10^{-3}$, $10^{-3}$, 0, 1, 1)}
\smallskip
\smallskip
\begin{tabular}{|c|c|c|}
\hline
\ \ \ & Time (minutes) & Memory  \\ 
\hline
Present approach  & 28 &  13GB  \\

Perturbation method  & 112 & 7.6GB \\

NS11 approach & 33 & 226MB\\
\hline
\end{tabular}

\label{table-1}
\end{centering}
\end{table}
Table~\ref{table-1} shows a comparison of computing resources required by three numerical 
methods. Compared to the perturbation method that requires large computing time 
(112 minutes to obtain a solution), the approximate method of NS11, and the present method 
are less expensive. In spite of being inconsistent, the approximate method of NS11 requires 
far less computing memory compared to the other two methods. This is because 
in NS11 the polarized transfer equation is solved in a azimuth independent Fourier 
basis, which thereby avoids introducing azimuth angle grids. The present approach 
requires larger memory because we now need to discretize the azimuth angle 
$\chi^\prime$ and store the huge matrix 
$\widetilde {{\mathcal{R}}}^{(k)KK^\prime}_{QQ^{\prime\prime}}
(m,x,x^\prime,\theta,{\bm\Omega^\prime},{\bm B})$. The memory requirement 
of the present approach is even larger than that of perturbation method 
because the former involves solving 54 coupled integral equations in Fourier 
basis, while the later involves solving 3 coupled integral equations in Stokes 
basis. However, unlike the present approach the convergence is not always 
guaranteed in the perturbation method, as angle-frequency coupling is more 
intricate in Stokes basis than in Fourier basis.
\subsection{The vertical field Hanle effect}
\label{vh}
\begin{figure*}
\centering
\includegraphics[height=12.0cm,width=8.5cm]{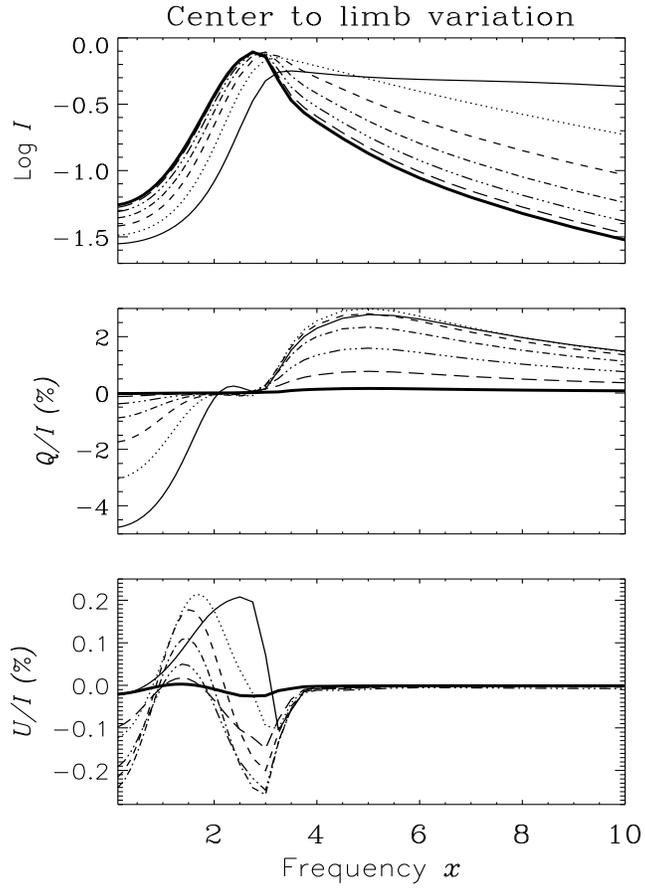}
\caption{Stokes $I$ and ratios $Q/I$ and  $U/I$ for different emergent 
angles ($\theta = \cos^{-1}\mu$) computed for a vertical magnetic field. The atmospheric model 
is same as in Fig.~\ref{comp-vh}. Different line types are: 
solid line - $\mu$ = 0.025, dotted line - $\mu$ = 0.129,  
dashed line - $\mu$ = 0.297, dot-dashed line - $\mu$ = 0.50, 
dash-triple-dotted line - $\mu$ =0.702, long-dashed line - $\mu$ = 0.871, and 
thick solid line - $\mu$ = 0.974.}
\label{comp-mu-vh}
\end{figure*}

It is expected that when the magnetic field is parallel to the symmetry axis 
of the slab (the atmospheric normal), the Hanle effect should vanish. In other words 
the Stokes $U$ parameter should be zero in this case. This characteristic 
behavior is satisfied when we work with angle-averaged redistribution functions. 
When angle-dependent redistribution function is used this behavior is not 
satisfied. In other words, Stokes $U$ does not vanish, in spite of the field 
being vertical, as long as the AD redistribution function is used. 
The non-zero emergent Stokes $U$ is due to coupling of Stokes $U$ 
to Stokes $I$ through the components 
$\widetilde {{\mathcal{R}}}^{(k)KK^\prime}_{QQ^{\prime}}$ for $k\neq0$. The reason 
for this unexpected behavior is also discussed by \citet{hf01} and numerically 
demonstrated in \citet{knnetal02}. We revisit this interesting problem in 
Figs.~\ref{comp-vh} and ~\ref{comp-mu-vh}. 
In Fig.~\ref{comp-vh} we show the Stokes profiles computed using NS11 and 
the present approach, and compare it with the results from 
perturbation method. The 
differences between the NS11 and the present approach are 
drastic in Stokes $U$ 
parameter. The Stokes $U$ in the frequency range $3 \lesssim x \lesssim 5$ 
has opposite signs. 
The present approach produces 
Stokes $U$ consistent with the perturbation method. 
The NS11 approach for approximation {\rm II} seems to be 
inadequate in computing $U$ in this particular problem. Such a large difference 
between the results obtained from NS11 approach and the perturbation method 
prompts us to conclude that it is safer 
to use AD frequency domains to solve AD Hanle transfer problems.\\
\indent Fig.~\ref{comp-mu-vh} shows the center to limb variation of linear 
polarization for a vertical magnetic field. The intensity exhibits the 
characteristic limb darkening in the line core and limb brightening 
in the wings. The $Q/I$ shows limb brightening throughout the line profile. 
The dependence on $\mu$ is non-monotonic 
in the core region in $U/I$. Since the angle-dependent 
functions become azimuthally symmetric at the disk center the $U/I$ 
approaches zero as the line of sight approaches the disk center. 
In the line wings, $U/I$ tend to zero for all $\mu$'s which is due to the 
Rayleigh scattering in the line wings, that produces $U/I=0$ by axisymmetry. 
\subsection{The Hanle effect with micro-turbulent magnetic field}
\label{micro-turbulent}
\begin{figure*}
\centering
\includegraphics[height=7.5cm,width=7.5cm]{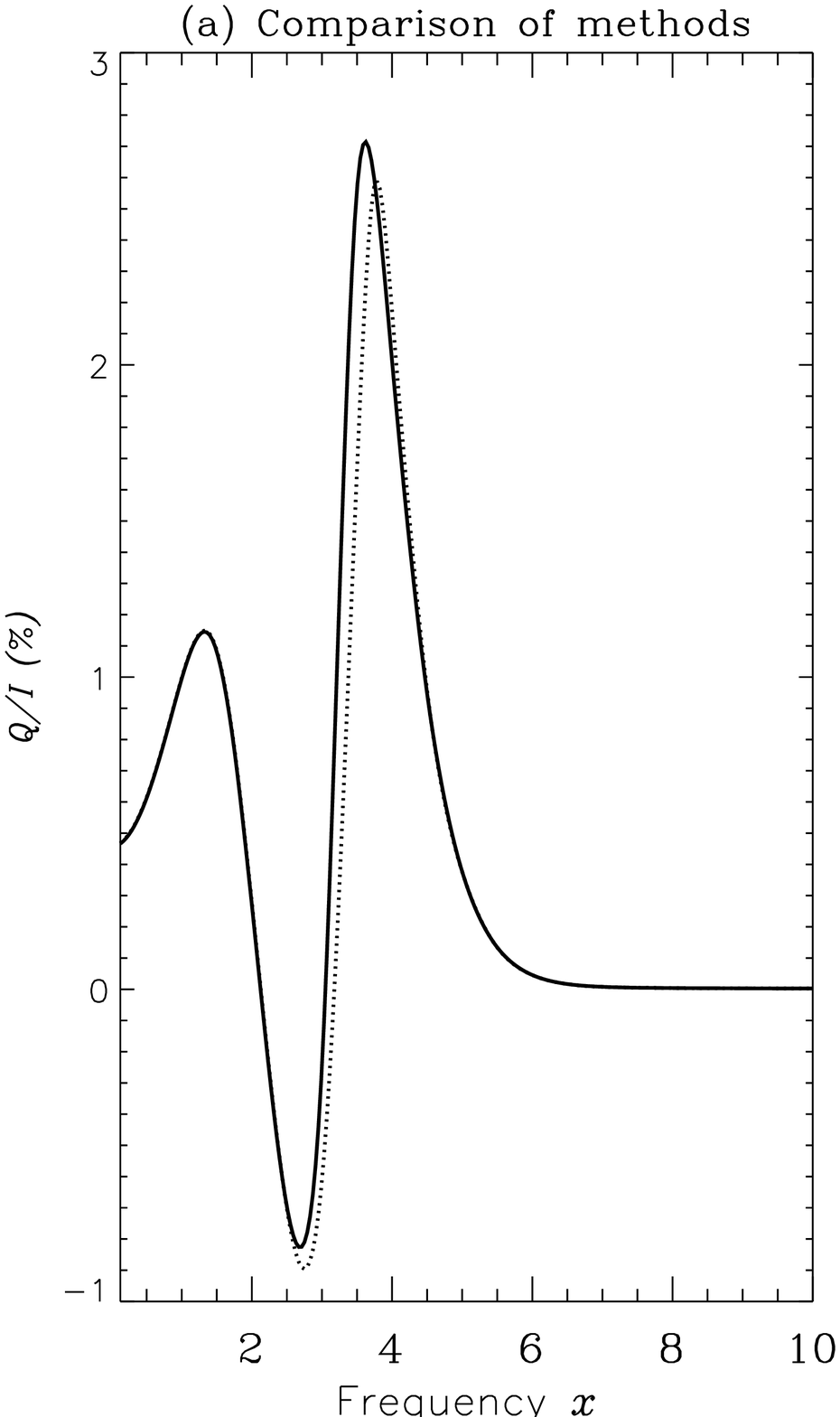}
\includegraphics[height=7.5cm,width=7.5cm]{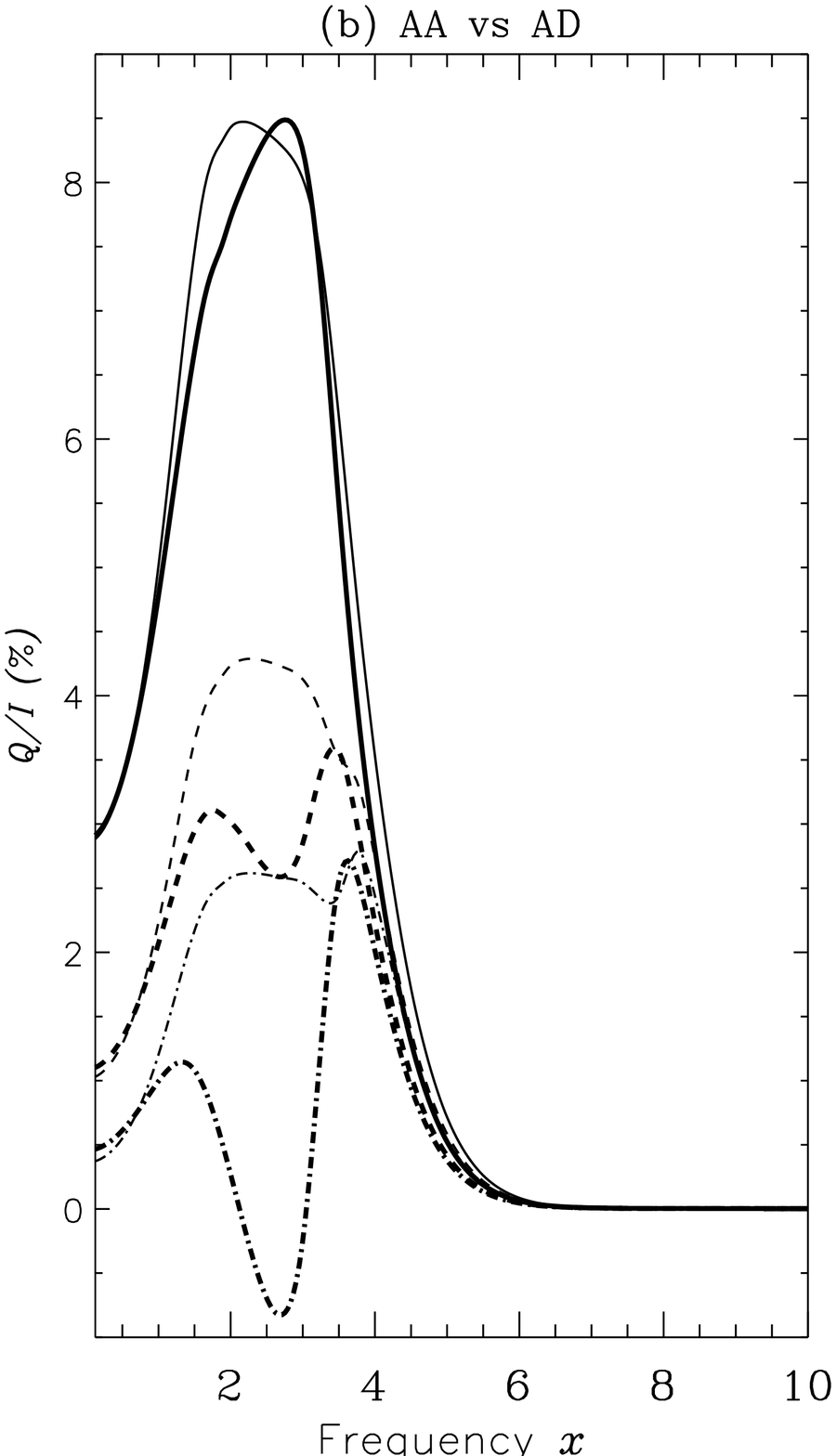}
\caption{Stokes $Q/I$ for emergent 
angle $\mu=0.112$ computed for the case of micro-turbulent magnetic field. 
The atmospheric model is 
($T$, $a$, $\epsilon$, $r$, $B_{\nu_0}$, $\Gamma_{E}/\Gamma_{R}$) = 
($10$, $10^{-3}$, $10^{-3}$, 0, 1, 0). Panel (a) shows the comparison of the 
present approach (solid line) and NS11 approach (dotted line). 
Panel (b) shows the comparison of angle-averaged and angle-dependent
cases. Different line types are : non-magnetic case - solid lines;
deterministic
magnetic field - dashed lines; and micro-turbulent magnetic field -
dot-dashed lines. The thick and thin lines represent angle-averaged and
angle-dependent cases respectively. For the micro-turbulent magnetic field 
case, $\Gamma = 1$ and for the deterministic magnetic field case, 
($\Gamma$, $\theta_B$, $\chi_B$) = 
(1, 30{\textdegree}, 0{\textdegree}).}
\label{comp-aa-ad-mt}
\end{figure*}
It is known that the presence of a weak turbulent magnetic field in the 
solar atmosphere can be 
detected using Hanle effect. In the case of CRD, \citet{hfetal09} showed 
that the polarization obtained using Hanle effect is quite sensitive to the 
choice of field strength distribution and in general, micro-turbulence is 
a safe approximation to represent weak turbulent magnetic fields. 
For a micro-turbulent magnetic field, the scale of variation of the field 
is small compared to the mean free path of the photons, 
and this allows to replace all the field dependent physical 
parameters by their averages over the magnetic field vector probability 
density function (PDF). In our problem this condition leads to 
the averaging of the 
magnetic kernel ${\mathcal N}^{r K}_{QQ^\prime}(m,{\bm B})$ over the magnetic 
field vector PDF. In this paper we use a PDF corresponding to the 
isotropic distribution of field orientation ($\theta_B$, $\chi_B$) and 
a single value of the field strength. As shown in 
\citet{stenflo82, stenflo94} the Hanle problem 
with this choice of PDF reduces then to a resonance polarization problem, 
with a modified value of $Q/I$. In other words, the micro-turbulent averaged magnetic 
kernel namely $<{\mathcal N}^{r K}_{QQ^\prime}(m,{\bm B})>$ becomes 
diagonal, and only $<{\mathcal N}^{r 2}_{00}(m,{\bm B})>$ element is of relevance. 
The explicit form of $<{\mathcal N}^{r 2}_{00}(m,{\bm B})>$ is given by 
(see Appendix B of \citet{hfetal09})
\begin{equation}
<{\mathcal N}^{r 2}_{00}(m,{\bm B})> \, = 1 - {2\over{5}}
{\Gamma^2(m)\over{1+\Gamma^2(m)}} - {2\over{5}} 
{4\Gamma^2(m)\over{1+4\Gamma^2(m)}},
\label{N2_00}
\end{equation}
where $\Gamma(m)$ denotes the Hanle $\Gamma$-parameter in different frequency 
domains. We refer the reader to Eqs. (89) of \citet{vb97} for the explicit form of 
$\Gamma(m)$ in different frequency domains where Hanle effect is operative. In 
frequency domains where Rayleigh scattering is present, $\Gamma(m)=0$. 
Fig.~\ref{comp-aa-ad-mt} (a) shows the ratio $Q/I$ obtained using the present approach 
and the NS11 approach in the presence of 
micro-turbulent magnetic field. The difference between the results obtained using 
these different methods mainly exists in the transition region $3 \lesssim x \lesssim 5$. 
Fig.~\ref{comp-aa-ad-mt} (b) shows the comparison of $Q/I$ profiles computed using 
angle-averaged and angle-dependent redistribution matrices. For AA computations, 
we use redistribution matrices computed in AA domains (approximation {\rm III} of 
\citet{vb97}), and for AD computations the 
corresponding AD redistribution matrices are computed in AD domains (approximation {\rm II} of 
\citet{vb97}). Three 
different sets of results are presented in Fig.~\ref{comp-aa-ad-mt} (b), namely 
non-magnetic, deterministic field, and micro-turbulent field results. 
We see that in the case of Hanle effect 
with micro-turbulent magnetic field, there is depolarization in the line core and in 
the near wing frequencies as compared to the other two cases. Our studies show that 
in the case of 
Hanle effect with micro-turbulent magnetic field, the differences 
between angle-averaged and angle-dependent results are prominent in thin slab cases 
and reduce considerably in thick slab cases. These differences between AA and AD results, 
especially in the $3 \lesssim x \lesssim 5$ region (apart from the line core), were already noticed by 
\citet{knnetal02}, for the deterministic magnetic field case. It is interesting to note 
that such differences get enhanced in the presence of a micro-turbulent magnetic field. This is 
probably because of the localization of line photons within the micro-turbulent 
scattering eddies, resulting in a relatively larger number of scattering. These differences 
are present both in the CRD scattering mechanism (see \citet{hfetal09}), as well as 
PRD as can be seen from Fig.~\ref{comp-aa-ad-mt} (b).

\section{Conclusions}
\label{conclusion}
We have solved the angle-dependent Hanle scattering problem using 
angle-dependent partial frequency redistribution (PRD) theory 
(approximation \rm{II} of \citet{vb97}). This 
computationally expensive problem is solved using 
an iterative method based on the Neumann series expansion (SEM). 
Following \citet{lsaknn11a} we decompose the Stokes parameters 
in terms of azimuthally symmetric Fourier coefficients, by expanding 
the Hanle redistribution matrix in terms of the radiation azimuth $\chi$. 
Only such a decomposition allows the use of angle-dependent frequency 
domains for solving angle-dependent Hanle scattering problems. 
In contrast a decomposition, based on the expansion in 
terms of $(\chi - \chi^\prime)$, as done in NS11, 
does not allow the use of angle-dependent domains to solve 
angle-dependent Hanle scattering problem. For this reason a simpler 
approach was suggested 
by NS11 which used 
angle-averaged frequency domains (approximation \rm{III} of \citet{vb97}), 
to solve 
angle-dependent Hanle transfer problem. We show that their approach does not 
always hold good. We have carried out a numerical study to show the 
differences between the solutions 
obtained by NS11 approach and the self-consistent approach used
now in this paper. The $U/I$ profiles in particular show significant differences 
in the core to wing transition region ($3 \lesssim x \lesssim 5$) of the line. 
The special 
case of vertical field Hanle effect is considered as a case study, 
and the differences 
between the NS11 and the present approach are examined. It is shown that the present 
method offers a self-consistent and accurate method of solving the difficult 
problem of AD partial redistribution with Hanle scattering. The interesting 
behavior of $Q/I$ profiles in the presence of micro-turbulent magnetic fields is also examined. 
We show that the differences between angle-averaged and angle-dependent solutions are 
enhanced by the presence of a micro-turbulent field. The differences are 
noticed in both the line core and near wing regions ($3\lesssim x \lesssim5$).

\section*{Acknowledgments}

We are grateful to the Referee for very useful and constructive suggestions which 
helped to greatly improve the paper.

\bibliographystyle{model3-num-names}
     
\end{document}